\documentclass[final]{siamltex}
\usepackage{amsfonts,amsmath,graphicx}
\newtheorem{claim}{Claim}

\newcommand{\twothmref}[2]{Theorems~\ref{thm:#1} and \ref{thm:#2}}
\newcommand{\lemref}[1]{Lemma~\ref{lem:#1}}
\newcommand{\twolemref}[2]{Lemmata~\ref{lem:#1} and \ref{lem:#2}}
\newcommand{\thmref}[1]{Theorem~\ref{thm:#1}}
\newcommand{\figref}[1]{Fig.~\ref{fig:#1}}
\newcommand{\secref}[1]{\S\ref{sec:#1}}
\newcommand{\corref}[1]{Corollary~\ref{cor:#1}}
\newcommand{\eqnref}[1]{\eqref{eqn:#1}}
\newcommand{\seclabel}[1]{\label{sec:#1}}
\newcommand{\figlabel}[1]{\label{fig:#1}}
\newcommand{\thmlabel}[1]{\label{thm:#1}}
\newcommand{\lemlabel}[1]{\label{lem:#1}}
\newcommand{\corlabel}[1]{\label{cor:#1}}
\newcommand{\eqnlabel}[1]{\label{eqn:#1}}

\newcommand{\Figure}[4][htb]{
\begin{figure}[#1]
\begin{center}#3\end{center}
\caption{\figlabel{#2}#4}
\end{figure}}

\DeclareMathOperator{\polylog}{polylog}
\newcommand{\etal}{ \emph{et al.}~}
\newcommand{\maxm}[1]{\ensuremath{\protect\max\{#1\}}}
\newcommand{\minm}[1]{\ensuremath{\protect\min\{#1\}}}
\newcommand{\floor}[1]{\ensuremath{\protect\lfloor#1\rfloor}}
\newcommand{\half}{\ensuremath{\protect\tfrac{1}{2}}}
\newcommand{\ceil}[1]{\ensuremath{\protect\lceil#1\rceil}}
\newcommand{\Oh}[1]{\ensuremath{\protect\mathcal{O}(#1)}}
\newcommand{\e}{\ensuremath{\boldsymbol{e}}}
\newcommand{\paran}[1]{\textup{(}#1\textup{)}}
\newcommand{\NP}{\ensuremath{\mathcal{NP}}}
\newcommand{\G}{\ensuremath{\mathcal{G}}}
\newcommand{\III}{\ensuremath{\mathcal{I}}}
\newcommand{\SSS}{\ensuremath{\mathcal{S}}}
\newcommand{\F}{\ensuremath{\mathcal{F}}}
\newcommand{\cn}[1]{\ensuremath{\chi(#1)}}
\newcommand{\acn}[1]{\ensuremath{\chi_{\textup{a}}(#1)}}
\newcommand{\tn}[1]{\ensuremath{\textup{\textsf{tn}}(#1)}}
\newcommand{\qn}[1]{\ensuremath{\textup{\textsf{qn}}(#1)}}
\newcommand{\sn}[1]{\ensuremath{\textup{\textsf{sn}}(#1)}}
\newcommand{\bw}[1]{\ensuremath{\textup{\textsf{bw}}(#1)}}
\newcommand{\tpw}[1]{\ensuremath{\textup{\textsf{tpw}}(#1)}}
\newcommand{\pw}[1]{\ensuremath{\textup{\textsf{pw}}(#1)}}
\newcommand{\tw}[1]{\ensuremath{\textup{\textsf{tw}}(#1)}}

\begin{document}

\renewcommand{\thefootnote}{\fnsymbol{footnote}}

\title{Layout of Graphs with Bounded Tree-Width\,\thanks{Submitted October 14, 2002. Revised \today. Results in this paper were presented at the GD~'02 \cite{DMW-GD02}, FST~TCS~'02 \cite{Wood-FSTTCS02}, and WG~'03 \cite{DujWoo-WG03} conferences.}}

\author{
Vida Dujmovi{\'c}\,\footnotemark[3]\ \footnotemark[4]\ \footnotemark[2] \and 
Pat Morin\,\footnotemark[4]\ \footnotemark[2] \and 
David R. Wood\,\footnotemark[4]\ \footnotemark[5]\ \footnotemark[2]}

\footnotetext[3]{School of Computer Science, McGill University, Montr{\'e}al,
Canada. E-mail: \texttt{vida@cs.mcgill.ca}}

\footnotetext[4]{School of Computer Science, Carleton University, Ottawa,
Canada.\\ E-mail: \texttt{\{morin,davidw\}@scs.carleton.ca} }

\footnotetext[5]{Department of Applied Mathematics, Charles University, Prague,
Czech Republic. Research supported by COMBSTRU.}

\footnotetext[2]{Research supported by NSERC.}

\maketitle

\begin{abstract}  A \emph{queue layout}  of a graph consists of a total order of the vertices, and a partition of the edges into  \emph{queues}, such that no two edges in the same queue are nested. The minimum number of queues in a queue layout of a graph is its \emph{queue-number}. A \emph{three-dimensional \paran{straight-line grid} drawing} of a graph represents the vertices by points in $\mathbb{Z}^3$ and the edges by non-crossing line-segments. This paper contributes three main results:

(1) It is proved that the minimum volume of a certain type of three-dimensional drawing of a graph $G$ is  closely related to the queue-number of $G$. In particular, if $G$ is an $n$-vertex member of a proper minor-closed family of graphs (such as a planar graph), then $G$ has a $\Oh{1}\times \Oh{1}\times \Oh{n}$ drawing if and only if $G$ has \Oh{1} queue-number. 
 
 (2) It is proved that queue-number is bounded by tree-width, thus resolving an open problem due to Ganley and Heath (2001), and disproving a conjecture of Pemmaraju (1992). This result provides renewed hope for the positive resolution of a number of open problems in the theory of queue layouts. 

(3) It is proved that graphs of bounded tree-width have three-dimensional drawings with \Oh{n} volume. This is the most general family of graphs known to admit three-dimensional drawings with \Oh{n} volume.

The proofs depend upon our results regarding \emph{track layouts} and \emph{tree-partitions} of graphs, which may be of independent interest.
\end{abstract}



\pagestyle{myheadings}
\thispagestyle{plain}
\markboth{VIDA DUJMOVI\'C, PAT MORIN, AND DAVID R. WOOD}{LAYOUT OF GRAPHS WITH BOUNDED TREE-WIDTH}

\renewcommand{\thefootnote}{\arabic{footnote}}

%%%%%%%%%%%%%%%%%%%%%%%%%%%%%%%%%%%%%%%%%%%%%%%%%%%%%%%%%%%%%%%%%%%%%%%%%%%
\section{Introduction}
\seclabel{Introduction}
%%%%%%%%%%%%%%%%%%%%%%%%%%%%%%%%%%%%%%%%%%%%%%%%%%%%%%%%%%%%%%%%%%%%%%%%%%%

A \emph{queue layout} of a graph consists of a total order of the vertices, and a partition of the edges into  \emph{queues}, such that no two edges in the same queue are nested. The dual concept of a \emph{stack layout}, introduced by  Ollmann~\cite{Ollmann73} and commonly called a \emph{book embedding}, is defined similarly, except that no two edges in the same \emph{stack} may cross. The minimum number of queues (respectively, stacks) in a queue layout (stack layout) of a graph is its \emph{queue-number} (\emph{stack-number}).  Queue layouts have been extensively studied  \cite{EI71, Hasunuma-GD03, HLR-SJDM92, HR-SJC92, Pemmaraju-PhD, RM-COCOON95, SS00, Tarjan72a} with applications in parallel process scheduling, fault-tolerant processing, matrix computations,  and sorting networks (see \cite{Pemmaraju-PhD} for a survey). Queue layouts of directed acyclic graphs  \cite{BCLR-JPDC96, HP-SJC99, HPT-SJC99, Pemmaraju-PhD} and posets \cite{HP-SJDM97, Pemmaraju-PhD} have also been investigated. Our motivation for studying queue layouts is a connection with three-dimensional graph drawing.

Graph drawing is concerned with the automatic generation of aesthetically
pleasing geometric representations of graphs. Graph drawing in the plane is
well-studied (see \cite{DETT99, KaufmannWagner01}). Motivated by experimental
evidence suggesting that displaying a graph in three dimensions is better than
in two \cite{WF94, WF96}, and applications  including information visualisation
\cite{WF94}, VLSI circuit design \cite{LR86}, and software engineering
\cite{WHF93}, there is a growing body of research in three-dimensional graph
drawing. In this paper we study \emph{three-dimensional straight-line grid
drawings}, or \emph{three-dimensional drawings} for short. In this model,
vertices are positioned at grid-points in $\mathbb{Z}^3$, and edges are drawn
as straight line-segments with no crossings \cite{CS-IPL97, CELR-Algo96, Giacomo-GD03, dGLW-CCCG02, DM-GD03, FLW-JGAA03, Hasunuma-GD03, Poranen-00, PTT99}. We focus on the problem of producing three-dimensional drawings with small volume. Three-dimensional drawings with the vertices in $\mathbb{R}^3$ have also been studied \cite{EG95, GTV96, CGT96, BF95, CDL95, HE-ISAAC00, CT95, HEQL98, Hong-GD01, HE-Algo03, MRS95, Ostry96}. Aesthetic criteria besides volume that have been considered include symmetry \cite{Hong-GD01, HE-ISAAC00, HE-Algo03, HEQL98}, aspect ratio \cite{CGT96, GTV96}, angular resolution \cite{GTV96,CGT96}, edge-separation \cite{CGT96, GTV96}, and convexity \cite{CDL95, CGT96, EG95, Steinitz22}.

The first main result of this paper reduces the question of whether a graph has a three-dimensional drawing with small volume to a question regarding queue layouts (\thmref{DrawingIFFQueue}). In particular, we prove that every $n$-vertex graph from a proper minor-closed graph family \G\ has a $\Oh{1}\times \Oh{1}\times \Oh{n}$ drawing if and only if \G\ has a \Oh{1} queue-number, and this result holds true when replacing \Oh{1} by $\Oh{\polylog n}$. Consider the family of planar graphs, which are minor-closed. (In the conference version of their paper) Felsner\etal\cite{FLW-JGAA03} asked whether every planar graph has a three-dimensional drawing with \Oh{n} volume?   Heath~\etal~\cite{HR-SJC92,HLR-SJDM92} asked whether every planar graph has \Oh{1} queue-number? By our result, these two open problems are almost equivalent in the following sense.  If every planar graph has \Oh{1} queue-number, then every planar graph has a three-dimensional drawing with \Oh{n} volume. Conversely, if every planar graph has a $\Oh{1}\times \Oh{1}\times \Oh{n}$ drawing, then every planar graph has \Oh{1} queue-number. It is possible, however, that planar graphs have unbounded queue-number, yet have say $\Oh{n^{1/3}}\times\Oh{n^{1/3}}\times\Oh{n^{1/3}}$ drawings.

Our other main results regard three-dimensional drawings and queue layouts of graphs with bounded tree-width. Tree-width, first defined by Halin~\cite{Halin76}, although largely unnoticed until independently rediscovered by Robertson and Seymour~\cite{RS-GraphMinors-II} and Arnborg and Proskurowski~\cite{AP-DAM89}, is a measure of the similarity of a graph to a tree (see \secref{IntroTreewidth} for the definition). Tree-width (or its special case, path-width) has been previously used in the context of graph drawing by Dujmovi\'{c}\etal\cite{Dujmovic-etal-ESA01}, Hlin\v{e}n\'{y}~\cite{Hlineny-JCTB03}, and Peng~\cite{Peng01}, for example.

The second main result is that the queue-number of a graph is bounded by its tree-width (\corref{TreewidthQueuenumber}). This solves an open problem due to Ganley and Heath~\cite{GH-DAM01}, who proved that stack-number is bounded by tree-width, and asked whether a similar relationship holds for queue-number. This result has significant implications for the above open problem (does every planar graph have \Oh{1} queue-number), and the  more general question (since planar graphs have stack-number at most four \cite{Yannakakis86}) of whether queue-number is bounded by stack-number. Heath~\etal~\cite{HR-SJC92,HLR-SJDM92} originally conjectured that both of these questions have an affirmative answer. More recently however, Pemmaraju~\cite{Pemmaraju-PhD} conjectured  that  the `stellated $K_3$', a planar $3$-tree, has $\Theta(\log n)$ queue-number, and provided evidence to support this conjecture (also see \cite{GH-DAM01}). This suggested that the answer to both of the above questions was negative. In particular, Pemmaraju~\cite{Pemmaraju-PhD} and Heath [private communication, 2002] conjectured that planar graphs have \Oh{\log n} queue-number. However, our result provides a queue-layout of \emph{any} $3$-tree, and thus the stellated $K_3$, with \Oh{1} queues. Hence our result disproves the first conjecture of Pemmaraju~\cite{Pemmaraju-PhD} mentioned above, and renews hope in an affirmative answer to the above open problems. 

The third main result is that every graph of bounded tree-width has a three-dimensional drawing with \Oh{n} volume.  The family of graphs of bounded tree-width includes most of the graphs previously known to admit  three-dimensional drawings with \Oh{n} volume (for example, outerplanar graphs), and also includes many graph families for which the previous best volume bound was \Oh{n^2} (for example, series-parallel graphs). Many graphs arising in applications of graph drawing do have small tree-width. Outerplanar and series-parallel graphs are the obvious examples. Another example arises in software engineering applications. Thorup~\cite{Thorup-IC98} proved that  the control-flow graphs of go-to free programs in many programming languages have tree-width bounded by a small constant; in particular, $3$ for Pascal and $6$ for C. Other families of graphs having bounded tree-width (for constant $k$) include: almost trees with parameter $k$, graphs with a feedback vertex set of size $k$, band-width $k$ graphs, cut-width $k$ graphs, planar graphs of radius $k$, and $k$-outerplanar graphs. If the size of a maximum clique is a constant $k$ then chordal, interval and circular arc graphs also have bounded tree-width. Thus, by our result,  all of these graphs have three-dimensional drawings with \Oh{n} volume, and \Oh{1} queue-number.

To prove our results for graphs of bounded tree-width, we employ a related structure called a tree-partition, introduced independently by Seese~\cite{Seese85} and Halin~\cite{Halin91}. A \emph{tree-partition} of a graph is a partition of its vertices into `bags' such that contracting each bag to a single vertex gives a forest (after deleting loops and replacing parallel edges by a single edge). In a result of independent interest, we prove that every $k$-tree has a tree-partition such that each bag induces a connected $(k-1)$-tree, amongst other properties. The second tool that we use is a \emph{track layout}, which consists of a vertex-colouring and a total order of each colour class, such that between any two colour classes no two edges cross.

\medskip The remainder of the paper is organised as follows. In \secref{Background} we introduce the required background material, and state our results regarding three-dimensional drawings and queue layouts, and compare these with results in the literature. In \secref{BasicTrackLayouts} we establish a number of results concerning track layouts.  That three-dimensional drawings and queue-layouts are closely related stems from the fact that three-dimensional drawings and queue layouts are both closely related to track layouts, as proved in \secref{Drawings} and \secref{QueueLayouts}, respectively. In \secref{TreePartitions} we prove the above-mentioned theorem for tree-partitions of $k$-trees, which is used in \secref{TreewidthTrackLayout} to construct track layouts of graphs with bounded tree-width. We conclude in \secref{Conclusion} with a number of open problems.

%%%%%%%%%%%%%%%%%%%%%%%%%%%%%%%%%%%%%%%%%%%%%%%%%%%%%%%%%%%%%%%%%%%%%%%%%%%
\section{Background and Results}
\seclabel{Background}
%%%%%%%%%%%%%%%%%%%%%%%%%%%%%%%%%%%%%%%%%%%%%%%%%%%%%%%%%%%%%%%%%%%%%%%%%%%

Throughout this paper all graphs $G$ are undirected, simple, and finite with vertex set $V(G)$ and edge set $E(G)$. The number of vertices and the maximum degree of $G$ are respectively denoted by $n=|V(G)|$ and $\Delta(G)$.  The subgraph induced by a set of vertices $A\subseteq V(G)$ is denoted by $G[A]$. For all disjoint subsets $A,B\subseteq V(G)$, the bipartite subgraph of $G$ with vertex set $A\cup B$  and edge set $\{vw\in E(G):v\in A,w\in B\}$  is denoted by $G[A,B]$. 

A graph $H$ is a \emph{minor} of a graph $G$ if $H$ is isomorphic to a graph obtained from a subgraph of $G$ by contracting edges. A family of graphs closed
under taking minors is \emph{proper} if it is not the class of all graphs.

A \emph{graph parameter} is a function $\alpha$ that assigns to every graph $G$ a non-negative integer $\alpha(G)$. Let \G\ be a family of graphs.  By $\alpha(\G)$ we denote the function $f:\mathbb{N}\rightarrow\mathbb{N}$, where $f(n)$ is the maximum  of $\alpha(G)$, taken over all $n$-vertex graphs $G\in\G$. We say \G\ has \emph{bounded} $\alpha$ if $\alpha(\G)\in\Oh{1}$. A graph parameter $\alpha$ is \emph{bounded by} a graph parameter $\beta$ (for some graph family \G), if there exists a  function $g$ such that $\alpha(G)\leq g(\beta(G))$ for every graph $G$ (in \G). 

%%%%%%%%%%%%%%%%%%%%%%%%%%%%%%%%%%%%%%%%%%%%%%%%%%%%%%%%%%%%%%%%%%%%%%%%%%%
\subsection{Tree-Width}
\seclabel{IntroTreewidth}
%%%%%%%%%%%%%%%%%%%%%%%%%%%%%%%%%%%%%%%%%%%%%%%%%%%%%%%%%%%%%%%%%%%%%%%%%%%%%%%

Let $G$ be a graph and let $T$ be a tree. An element of $V(T)$ is called a \emph{node}.  Let $\{T_x\subseteq V(G):x\in V(T)\}$ be a  set of subsets of $V(G)$ indexed by the nodes of $T$. Each $T_x$ is called a \emph{bag}. The pair $(T,\{T_x:x\in V(T)\})$ is a \emph{tree-decomposition} of $G$ if:

\begin{remunerate} 

\item $\displaystyle \bigcup_{x\in V(T)}\!T_x\,=\,V(G)$ (that is, every vertex of $G$ is in at least one bag), 

\item $\forall$ edge $vw$ of $G$, $\exists$ node $x$ of $T$ such that  $v\in T_x$ and $w\in T_x$, and  

\item $\forall$ nodes $x,y,z$ of $T$, if $y$ is on the path from $x$ to $z$ in $T$, then $T_x\cap T_z\subseteq T_y$. 

\end{remunerate}

The \emph{width} of a tree-decomposition is one less than the maximum cardinality of a bag. A \emph{path-decomposition} is a tree-decomposition where the tree $T$ is a path $T=(x_1,x_2,\dots,x_m)$, which is simply identified by the sequence of bags $T_1,T_2,\dots,T_m$ where each $T_i=T_{x_i}$. The \emph{path-width} (respectively, \emph{tree-width}) of a graph $G$, denoted by \pw{G} (\tw{G}), is the minimum width of a path- (tree-) decomposition of $G$.  Graphs with tree-width at most one are precisely the forests.  Graphs with tree-width at most two are called \emph{series-parallel}\footnote{`Series-parallel digraphs' are often defined in terms of certain `series' and `parallel' composition operations. The underlying  undirected graph of such a digraph has tree-width at most two (see \cite{Bodlaender-TCS98}).}, and are characterised as those graphs with no $K_4$ minor (see \cite{Bodlaender-TCS98}). 

A \emph{$k$-tree} for some $k\in\mathbb{N}$ is defined recursively as follows. The empty graph is a $k$-tree, and the graph obtained from a $k$-tree by adding a new vertex adjacent to each vertex of a clique with at most $k$ vertices is also a $k$-tree. This definition of a $k$-tree is by  Reed~\cite{Reed-AlgoTreeWidth03}. The following more restrictive definition of a
$k$-tree, which we call `strict', was introduced by Arnborg and Proskurowski~\cite{AP-DAM89}, and is more often used in the literature.  A $k$-clique is a \emph{strict $k$-tree},
and the graph obtained from a strict $k$-tree by adding a new vertex adjacent
to each vertex of a $k$-clique is also a strict $k$-tree. Obviously the strict
$k$-trees are a proper sub-class of the $k$-trees.  A subgraph of a $k$-tree is
called a \emph{partial $k$-tree}, and a subgraph of a strict $k$-tree is called
a \emph{partial strict $k$-tree}. The following result is well known (see for
example \cite{Bodlaender-TCS98,Reed-AlgoTreeWidth03}). A \emph{chord} of a cycle $C$ is an edge not in $C$ whose end-vertices are both in $C$. A graph is \emph{chordal} if every cycle on at least four vertices has a chord.

\begin{lemma}
\lemlabel{TreewidthCharacterisation}
Let $G$ be a graph. The following are equivalent:
\begin{remunerate}
\item $G$ has tree-width $\tw{G}\leq k$,
\item $G$ is a partial $k$-tree,
\item $G$ is a partial strict $k$-tree,
\item $G$ is a subgraph of a chordal graph that has no clique on
$k+2$ vertices.
\end{remunerate}
\end{lemma}

\begin{proof} 
Scheffler~\cite{Scheffler89} proved that (1) and (3) are equivalent. That (1) and (4) are equivalent is due to Robertson and Seymour~\cite{RS-GraphMinors-II}. That (2) and (4) are equivalent is the characterisation of chordal graphs in terms of `perfect elimination' vertex-orderings due to Fulkerson and Gross~\cite{FG65}.  
\hfill\end{proof}

%%%%%%%%%%%%%%%%%%%%%%%%%%%%%%%%%%%%%%%%%%%%%%%%%%%%%%%%%%%%%%%%%%%%%%%%%%%
\subsection{Tree-Partitions}
\seclabel{IntroTreePartition}
%%%%%%%%%%%%%%%%%%%%%%%%%%%%%%%%%%%%%%%%%%%%%%%%%%%%%%%%%%%%%%%%%%%%%%%%%%%

As in the definition of a tree-decomposition, let $G$ be  graph and let
$\{T_x\subseteq V(G):x\in V(T)\}$ be a set of subsets of $V(G)$ (called
\emph{bags}) indexed by the nodes of a tree $T$.  The pair $(T,\{T_x:x\in
V(T)\})$ is a \emph{tree-partition} of $G$ if  
\begin{remunerate}
\item $\forall$ distinct nodes $x$ and $y$ of $T$, $T_x\cap T_y=\emptyset$, and
\item $\forall$ edge $vw$ of $G$,  either 
\begin{romannum}
\item $\exists$ node $x$ of $T$ with $v\in T_x$ and $w\in T_x$
($vw$ is called an \emph{intra-bag} edge), or 
\item $\exists$ edge $xy$ of $T$ with $v\in T_x$ and $w\in T_y$
($vw$ is called an \emph{inter-bag} edge).
\end{romannum}
\end{remunerate}

The main property of tree-partitions that has been studied in the literature 
is the maximum cardinality of a bag, called the \emph{width} of the tree-partition \cite{BodEng-JAlg97,Halin91,Seese85,DO-JGT95,DO-DM96}. The minimum width over all tree-partitions of a graph $G$ is the
\emph{tree-partition-width}\footnote{Tree-partition-width has also been called
\emph{strong tree-width} \cite{Seese85,BodEng-JAlg97}.} of $G$, denoted by $\tpw{G}$. A graph with bounded degree has bounded tree-partition-width if and only if it has bounded tree-width \cite{DO-DM96}. In particular, for every graph $G$,  Ding and Oporowski~\cite{DO-JGT95} proved that $\tpw{G}\leq24\,\tw{G}\Delta(G)$, and Seese~\cite{Seese85} proved that $\tw{G}\leq2\,\tpw{G}-1$.

\thmref{TreePartition} provides a tree-partition of a $k$-tree $G$ with additional features besides small width. First, the subgraph induced by each bag is a connected $(k-1)$-tree. This allows us to perform induction on $k$. Second, in each non-root bag $T_x$ the set of vertices in the parent bag of $x$ with a neighbour in $T_x$ form a clique. This feature is crucial in the intended application (\thmref{TreewidthTracknumber}). Finally the tree-partition has width at most $\max\{1,k(\Delta(G)-1)\}$, which represents a constant-factor improvement over the above result by Ding and Oporowski~\cite{DO-JGT95} in the case of $k$-trees.

%%%%%%%%%%%%%%%%%%%%%%%%%%%%%%%%%%%%%%%%%%%%%%%%%%%%%%%%%%%%%%%%%%%%%%%%%%%%%%
\subsection{Track Layouts}
\seclabel{IntroTrackLayouts}
%%%%%%%%%%%%%%%%%%%%%%%%%%%%%%%%%%%%%%%%%%%%%%%%%%%%%%%%%%%%%%%%%%%%%%%%%%%%%%

Let $G$ be a graph. A \emph{colouring} of $G$ is a partition $\{V_i:i\in I\}$ of $V(G)$, where $I$ is a set of \emph{colours}, such that for every edge $vw$ of $G$, if $v\in V_i$ and $w\in V_j$ then $i\ne j$.  Each set $V_i$ is called a \emph{colour class}.  A colouring of $G$ with $c$ colours is a \emph{$c$-colouring}, and we say that $G$ is \emph{$c$-colourable}. The \emph{chromatic number} of $G$, denoted by \cn{G}, is the minimum $c$ such that $G$ is $c$-colourable.

If $<_i$ is a total order of a colour class $V_i$, then we call the pair $(V_i,<_i)$ a \emph{track}.  If $\{V_i:i\in I\}$ is a colouring of $G$, and $(V_i,<_i)$ is a track, for each colour $i\in I$, then we say $\{(V_i,<_i):i\in I\}$ is a \emph{track assignment} of $G$ \emph{indexed by} $I$.   Note that at times it will be convenient to also refer to a colour $i\in I$ and the colour class $V_i$ as a \emph{track}. The precise  meaning will always be clear from the context.  A \emph{$t$-track assignment} is a track assignment with $t$ tracks. 

As illustrated in \figref{xCrossing}, an \emph{X-crossing} in a track assignment consists of two edges $vw$ and $xy$ such that $v<_ix$ and $y<_jw$, for distinct tracks $V_i$ and $V_j$. A $t$-track assignment with no X-crossing is called a \emph{$t$-track layout}.  The \emph{track-number} of a graph $G$, denoted by \tn{G}, is the minimum $t$ such that $G$ has a $t$-track layout. 

\Figure{xCrossing}{\includegraphics{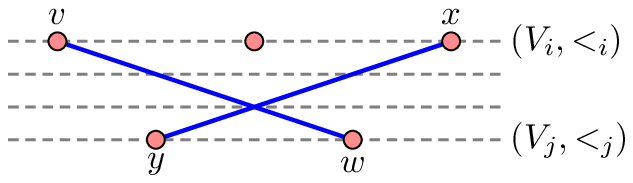}}{An example of an X-crossing in a track assignment.}

Let $\{(V_i,<_i):i\in I\}$ be a $t$-track layout of a graph $G$. The \emph{span} of an edge $vw$ of $G$, with respect to a numbering of the tracks $I=\{1,2,\dots,t\}$, is defined to be $|i-j|$ where $v\in V_i$ and $w\in V_j$. 

Track layouts will be central in most of our proofs.  To enable comparison of our results to those in the literature we now introduce the notion of an `improper' track layout. A \emph{improper colouring} of a graph $G$ is simply a partition $\{V_i:i\in I\}$ of $V(G)$. Here adjacent vertices may be in the same colour class. A track of an improper colouring is defined as above.  Suppose $\{V_i:i\in I\}$ is an improper colouring of $G$, and $(V_i,<_i)$ is a track, for each colour $i\in I$. An edge with both end-vertices in the same track is called an \emph{intra-track} edge; otherwise it is called an \emph{inter-track} edge. We say $\{(V_i,<_i):i\in I\}$ is an \emph{improper track assignment} of $G$ if, for all intra-track edges $vw\in E(G)$ with $v\in V_i$ and $w\in V_i$ for some $i\in I$, there is no vertex $x$ with $v<_ix<_iw$. That is, adjacent vertices in the same track are consecutive in that track. An improper $t$-track assignment with no X-crossing is called an \emph{improper $t$-track layout}\footnote{In \cite{DMW-GD02,DujWoo-TR-02-03,Wood-FSTTCS02} we called a track layout an \emph{ordered layering with no X-crossing and no intra-layer edges}, and an improper track layout was called an \emph{ordered layering with no X-crossing}.}. 

\begin{lemma}
\lemlabel{ImproperProper}
If a graph $G$ has an improper $t$-track layout, then $G$ has a $2t$-track layout.
\end{lemma}

\begin{proof} For every track $V_i$ of an improper $t$-track layout of $G$, let $V_i'$ be a new track. Move every second vertex from $V_i$ to $V_i'$, such that $V_i'$ inherits its total order from the original $V_i$. Clearly there is no intra-track edge and no X-crossing. Thus we obtain a $2t$-track layout of $G$.  \hfill\end{proof}

Hence the track-number of a graph is at most twice its `improper track-number'. The following lemma, which was jointly discovered with Giuseppe Liotta, gives a compelling reason to only consider proper track layouts. Similar ideas can be found in \cite{FLW-JGAA03,dGLW-CCCG02}. Let $vw$ be an edge of a graph $G$. Let $G'$ be the graph obtained from $G$ by adding a new vertex $x$ only adjacent to $v$ and $w$. We say $x$ is an \emph{ear}, and $G'$ is obtained from $G$ by \emph{adding an ear to} $vw$.

\begin{lemma}
\lemlabel{AddingEars}
Let \G\ be a class of graphs closed under the addition of ears \paran{for example, series-parallel graphs or planar graphs}. If every graph in \G\ has an improper $t$-track layout for some constant $t$, then every graph in \G\ has a \paran{proper} $t$-track layout.
\end{lemma}

\begin{proof} For any graph $G\in\G$, let $G'$ be the graph obtained from $G$ by adding $t$ ears to every edge of $G$. By assumption, $G'$ has an improper $t$-track layout. Suppose that there is an edge $vw$ of $G$ such that $v$ and $w$ are in the same track. None of the ears added to $vw$ are on the same track, as otherwise adjacent vertices would not be consecutive in that track. Thus there is a track containing at least two of the ears added to $vw$. However, this implies that there is an X-crossing, which is a contradiction. Thus the end-vertices of every edge of $G$ are in distinct tracks. Hence the improper $t$-track layout of $G'$ contains a $t$-track layout of $G$.
\hfill\end{proof}

\twolemref{ImproperProper}{AddingEars} imply that only for relatively small classes of graphs will the distinction between track layouts and improper track layouts be significant. We therefore chose to work with the less cumbersome notion of a track layout. The following theorem summarises our bounds on the track-number of a graph.

\begin{theorem}
\thmlabel{TrackLayoutSummary}
Let $G$ be a graph with maximum degree $\Delta(G)$,  path-width \pw{G}, tree-partition-width \tpw{G}, and tree-width \tw{G}. The track-number of  $G$ satisfies: 
\begin{remunerate}
\item[\textup{(a)}] $\tn{G}\,\leq\,\pw{G}+1\,\leq\,1+(\tw{G}+1)\,\log n$,
\item[\textup{(b)}] $\tn{G}\,\leq\,3\,\tpw{G}\,\leq\,72\,\Delta(G)\,\tw{G}$,
\item[\textup{(c)}] $\tn{G}\,\leq\,3^{\,\tw{G}}\cdot6^{(4^{\,\tw{G}}-3\,\tw{G}-1)/9}$.
\end{remunerate}
\end{theorem}

\begin{proof} 
Part (a) follows from \lemref{PathDecomp2TrackLayout}, and the fact that $\pw{G}\leq(\tw{G}+1)\log n$ (see \cite{Bodlaender-TCS98}). Note that $\tn{G}\leq1+(\tw{G}+1)\log n$ can be proved directly using a separator-based approach similar to that used to prove $\pw{G}\leq(\tw{G}+1)\log n$. Part (b) follows from \lemref{TreePartition2TrackLayout}  in \secref{BasicTrackLayouts}, and the result of Ding and Oporowski~\cite{DO-JGT95} discussed in \secref{IntroTreePartition}. Part (c) is \thmref{TreewidthTracknumber}.
\hfill\end{proof}

%%%%%%%%%%%%%%%%%%%%%%%%%%%%%%%%%%%%%%%%%%%%%%%%%%%%%%%%%%%%%%%%%%%%%%%%%%%%%%%
\subsection{Vertex-Orderings}
\seclabel{IntroVertexOrderings}
%%%%%%%%%%%%%%%%%%%%%%%%%%%%%%%%%%%%%%%%%%%%%%%%%%%%%%%%%%%%%%%%%%%%%%%%%%%%%%%

Let $G$ be a graph. A total order $\sigma=(v_1,v_2,\dots,v_n)$ of $V(G)$ is called a \emph{vertex-ordering} of $G$. Suppose $G$ is connected. The \emph{depth} of a vertex $v_i$ in $\sigma$ is the graph-theoretic distance between $v_1$ and $v_i$ in $G$. We say $\sigma$ is a \emph{breadth-first} vertex-ordering if for all vertices $v$ and $w$ with $v<_{\sigma}w$, the depth of $v$ in $\sigma$ is no more than the depth of $w$ in $\sigma$. Vertex-orderings, and in particular, vertex-orderings of trees will be used extensively in this paper.  Consider a breadth-first vertex-ordering $\sigma$ of a tree $T$ such that vertices at depth $d\geq1$ are ordered with respect to the ordering of vertices at depth $d-1$. In particular, if $v$ and $x$ are vertices at depth $d$ with respective parents $w$ and $y$ at depth $d-1$ with $w<_\sigma y$ then $v<_\sigma x$. Such a vertex-ordering is called a \emph{lexicographical} breadth-first vertex-ordering of $T$, and is illustrated in \figref{lexTree}.

\Figure{lexTree}{\includegraphics{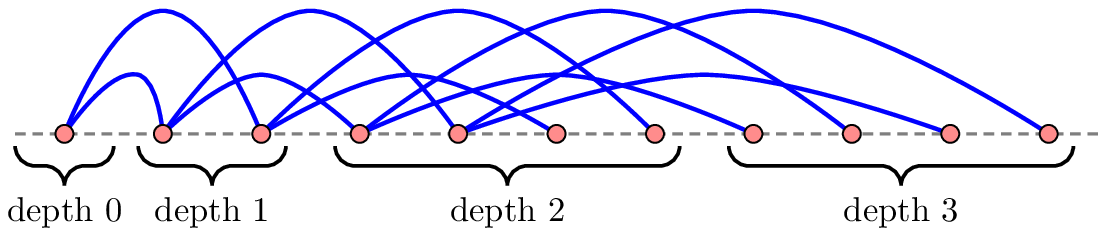}}{A lexicographical breadth-first vertex-ordering of a tree.}

%%%%%%%%%%%%%%%%%%%%%%%%%%%%%%%%%%%%%%%%%%%%%%%%%%%%%%%%%%%%%%%%%%%%
\subsection{Queue Layouts}
\seclabel{IntroQueues}
%%%%%%%%%%%%%%%%%%%%%%%%%%%%%%%%%%%%%%%%%%%%%%%%%%%%%%%%%%%%%%%%%%%%

A \emph{queue layout} of a graph $G$ consists of a vertex-ordering $\sigma$ of $G$, and a partition of $E(G)$ into \emph{queues}, such that no two edges in the same queue are \emph{nested} with respect to $\sigma$. That is, there are no edges $vw$ and $xy$ in a single queue with $v<_\sigma x<_\sigma y<_\sigma w$. 
The minimum number of queues in a queue layout of $G$ is called the \emph{queue-number} of $G$, and is denoted by $\qn{G}$. 
A similar concept is that of a \emph{stack layout} (or \emph{book embedding}), which  consists of a vertex-ordering $\sigma$ of $G$, and a partition of $E(G)$ into  \emph{stacks} (or \emph{pages}) such that there are no edges $vw$ and $xy$ in a single stack with $v<_\sigma x<_\sigma w<_\sigma y$.  The minimum number of stacks in a stack layout of $G$  is called the \emph{stack-number} (or \emph{page-number} or \emph{book-thickness}) of $G$, and is denoted by $\sn{G}$.   A queue (respectively, stack) layout with $k$ queues (stacks) is called a $k$-\emph{queue} ($k$-\emph{stack}) \emph{layout}, and a graph that admits a $k$-queue ($k$-stack) layout is called a \emph{$k$-queue} (\emph{$k$-stack}) \emph{graph}.

Heath and Rosenberg~\cite{HR-SJC92} characterised $1$-queue graphs as the `arched levelled planar' graphs, and proved that it is \NP-complete to recognise such graphs. This result is in contrast to the situation for stack layouts --- 1-stack graphs are precisely the outerplanar graphs \cite{BK79}, which can  be recognised in polynomial time. Heath\etal\cite{HLR-SJDM92} proved that 1-stack graphs are $2$-queue graphs (rediscovered by Rengarajan and Veni~Madhavan~\cite{RM-COCOON95}), and that $1$-queue graphs are $2$-stack graphs.

While it is \NP-hard to minimise the number of stacks in a stack layout given a fixed vertex-ordering \cite{GJMP80}, the analogous problem for queue layouts  can be solved as follows. A $k$-\emph{rainbow} in a vertex-ordering $\sigma$ consists of a matching $\{v_iw_i:1\leq i\leq k\}$ such that $v_1<_\sigma  v_2<_\sigma\dots<_\sigma  v_k<_\sigma w_k<_\sigma w_{k-1}<_\sigma\dots<_\sigma w_1$, as illustrated in \figref{Rainbow}.

\Figure{Rainbow}{\includegraphics{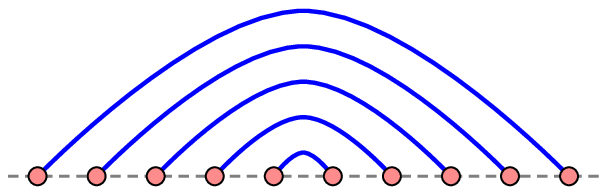}}{A rainbow of five edges in a
vertex-ordering.}

A vertex-ordering containing a $k$-rainbow needs at least $k$ queues.  A straightforward application of Dilworth's Theorem \cite{Dilworth50} proves the converse. That is, a fixed vertex-ordering admits a $k$-queue layout where $k$ is the size of the largest rainbow. (Heath and Rosenberg~\cite{HR-SJC92}  describe a $\Oh{m\log\log n}$ time algorithm to compute the queue assignment.)\ Thus determining $\qn{G}$ can be viewed as the following vertex-ordering problem.  

\begin{lemma}[\cite{HR-SJC92}]
\lemlabel{Rainbow}
The queue-number $\qn{G}$  of a graph $G$ is the minimum, taken over all vertex-orderings $\sigma$ of $G$, of the maximum  size of a rainbow in $\sigma$. \hfill\endproof
\end{lemma}

Stack and/or queue layouts of $k$-trees have previously been investigated in \cite{CLR87,RM-COCOON95,GH-DAM01}. A $1$-tree is a $1$-queue graph, since in a lexicographical breadth-first vertex-ordering of a tree no two edges are nested (see \figref{lexTree}). Chung\etal\cite{CLR87} proved that in a depth-first vertex-ordering of a tree no two edges cross. Thus $1$-trees are $1$-stack graphs. Rengarajan and Veni~Madhavan~\cite{RM-COCOON95} proved that graphs with tree-width at most two (the series parallel graphs) are $2$-stack and $3$-queue graphs\footnote{In \cite{DujWoo-TR-02-03} we give a simple proof based on \thmref{TreePartition} for the result by Rengarajan and Veni~Madhavan~\cite{RM-COCOON95} that every series-parallel graph has a $3$-queue layout.}. Improper track layouts are implicit in the work of Heath\etal\cite{HLR-SJDM92} and Rengarajan and Veni~Madhavan~\cite{RM-COCOON95}. In \secref{QueueLayouts} we prove the following fundamental relationship  between queue and track layouts.

\begin{theorem} 
\thmlabel{QueuenumberIFFTracknumber}  
For every graph $G$, $\qn{G}\leq\tn{G}-1$. Moreover, if \G\ is any proper minor-closed graph family, then \G\ has queue-number $\qn{\G}\in\F(n)$ if and only if \G\ has track-number $\tn{\G}\in\F(n)$, where $\F(n)$ is any family of functions closed under multiplication \paran{such as \Oh{1} or $\Oh{\polylog n}$}. 
\end{theorem} 

Ganley and Heath~\cite{GH-DAM01} proved that every graph $G$ has stack-number $\sn{G}\leq\tw{G}+1$ (using a depth-first traversal of a tree-decomposition), and asked whether queue-number is bounded by tree-width? One of the principal results of this paper is to solve this question in the affirmative. Applying \twothmref{TrackLayoutSummary}{QueuenumberIFFTracknumber} we have the following.

\begin{theorem}
\thmlabel{QueuenumberResults}
Let $G$ be a graph with maximum degree $\Delta(G)$,  path-width \pw{G}, tree-partition-width \tpw{G}, and tree-width \tw{G}.  The queue-number $\qn{G}$ satisfies\footnote{In \cite{Wood-FSTTCS02}  we obtained an alternative proof that $\qn{G}\leq\pw{G}$ using the `vertex separation number' of a graph (which equals its  path-width), and applying \lemref{Rainbow} directly we proved that  $\qn{G}\leq\frac{3}{2}\,\tpw{G}$, and thus $\qn{G}\leq36\,\Delta(G)\,\tw{G}$.}:
\begin{remunerate}
\item[\textup{(a)}] $\qn{G}\,\leq\,\pw{G}\,\leq\,(\tw{G}+1)\,\log n$,
\item[\textup{(b)}] $\qn{G}\,\leq\,3\,\tpw{G}-1\,\leq\,72\Delta(G)\,\tw{G}-1$,
\item[\textup{(c)}] $\qn{G}\,\leq\,3^{\,\tw{G}}\cdot6^{(4^{\,\tw{G}}-3\,\tw{G}-1)/9}-1$.
\hfill\endproof
\end{remunerate}
\end{theorem}

A similar upper bound to \thmref{QueuenumberResults}(a) is obtained by Heath and Rosenberg~\cite{HR-SJC92}, who proved that every graph $G$ has $\qn{G}\leq\ceil{\half\bw{G}}$, where $\bw{G}$ is the band-width of $G$. In many cases this result is weaker than \thmref{QueuenumberResults}(a)  since $\pw{G}\leq\bw{G}$ (see \cite{DPS-GraphLayouts}). More importantly,  we have the following corollary of \thmref{QueuenumberResults}(c).

\begin{corollary}
\corlabel{TreewidthQueuenumber}
Queue-number is bounded by tree-width, and hence graphs with bounded tree-width have bounded queue-number.\hfill\endproof
\end{corollary}

%%%%%%%%%%%%%%%%%%%%%%%%%%%%%%%%%%%%%%%%%%%%%%%%%%%%%%%%%%%%%%%%%%%%
\subsection{Three-Dimensional Drawings}
\seclabel{IntroDrawings}
%%%%%%%%%%%%%%%%%%%%%%%%%%%%%%%%%%%%%%%%%%%%%%%%%%%%%%%%%%%%%%%%%%%%

A \emph{three-dimensional straight-line grid drawing} of a graph, henceforth called a \emph{three-dimensional drawing}, represents the vertices by distinct points in $\mathbb{Z}^3$ (called \emph{grid-points}), and represents each edge as a line-segment between its end-vertices, such that edges only intersect at common end-vertices, and an edge only intersects a vertex that is an end-vertex of that edge. 

In contrast to the case in the plane, a folklore result states that every graph has a three-dimensional drawing. Such a drawing can be constructed using the `moment curve' algorithm in which vertex $v_i$, $1\leq i\leq n$, is represented by the grid-point  $(i,i^2,i^3)$. It is easily seen --- compare with \lemref{TrackLayout2Drawing} --- that no two edges cross. (Two edges \emph{cross} if they intersect at some point other than a common end-vertex.)

Since every graph has a three-dimensional drawing, we are interested in optimising certain measures of the aesthetic quality of a drawing. If a three-dimensional drawing is contained in an axis-aligned box with side lengths $X-1$, $Y-1$ and $Z-1$, then  we speak of an $X\times Y\times Z$ drawing with \emph{volume} $X\cdot Y\cdot Z$ and \emph{aspect ratio} $\maxm{X,Y,Z}/\minm{X,Y,Z}$.  This paper considers the problem of producing a three-dimensional drawing of a given graph with small volume, and with small aspect ratio as a secondary criterion.

Observe that the drawings produced by the moment curve algorithm have $\Oh{n^6}$ volume. Cohen\etal\cite{CELR-Algo96} improved this bound, by proving that  if $p$ is a prime with $n< p\leq 2n$, and each vertex $v_i$ is represented by the grid-point  $(i,i^2\bmod{p},i^3\bmod{p})$, then there is still no crossing. This construction is a generalisation of an analogous two-dimensional technique due to Erd\H{o}s~\cite{Erdos51}. Furthermore, Cohen\etal\cite{CELR-Algo96} proved that  the resulting  \Oh{n^3} volume bound is asymptotically optimal in the case of the complete graph $K_n$. It is therefore of interest to identify fixed graph parameters  that allow for three-dimensional drawings with small volume.  

The first such parameter to be studied was the chromatic number  \cite{CS-IPL97,PTT99}. Calamoneri and Sterbini~\cite{CS-IPL97} proved that every $4$-colourable graph has a three-dimensional drawing with \Oh{n^2} volume. Generalising this result, Pach\etal\cite{PTT99} proved that graphs of bounded chromatic number have three-dimensional drawings with \Oh{n^2} volume, and that this bound is asymptotically optimal for the complete bipartite graph with equal sized bipartitions. If $p$ is a suitably chosen prime, the main step of this algorithm represents the vertices in the $i$th colour class by grid-points in the set  $\{(i,t,it):t\equiv i^2\!\!\pmod{p}\}$. It follows that the volume bound is $\Oh{k^2n^2}$ for $k$-colourable graphs.

The lower bound of Pach\etal\cite{PTT99} for the complete bipartite graph was generalised by Bose\etal\cite{BCMW-JGAA} for all graphs. They proved that every three-dimensional drawing with $n$ vertices and $m$ edges has  volume at least $\frac{1}{8}(n+m)$. In particular, the maximum number of edges in an $X\times Y\times Z$ drawing is  exactly $(2X-1)(2Y-1)(2Z-1)-XYZ$. For example, graphs admitting three-dimensional drawings with \Oh{n} volume have \Oh{n} edges.   

The first non-trivial \Oh{n} volume bound was established by Felsner\etal\cite{FLW-JGAA03} for outerplanar graphs. Their elegant algorithm  `wraps' a two-dimensional drawing around a triangular prism to obtain an improper $3$-track layout (see \twolemref{TreeTrackLayout}{Rolling} for more on this method). Poranen~\cite{Poranen-00} proved that series-parallel digraphs have upward three-dimensional drawings with \Oh{n^3} volume, and that this bound can be improved to \Oh{n^2} and \Oh{n} in certain special cases. Di~Giacomo~\cite{dGLW-CCCG02} proved that series-parallel graphs with maximum degree three have three-dimensional drawings with \Oh{n} volume.

In \secref{Drawings} we prove the following intrinsic relationship between three-dimensional drawings and track layouts.

\begin{theorem}
\thmlabel{DrawingIFFTracknumber}
Every graph $G$ has a $\Oh{\tn{G}}\times\Oh{\tn{G}}\times\Oh{n}$ drawing. Moreover,  $G$ has a $\F(n)\times\F(n)\times \Oh{n}$ drawing if and only if $G$ has track-number $\tn{G}\in\F(n)$, where  $\F(n)$ is a family of functions closed under multiplication.
\end{theorem}

Of course, every graph has an $n$-track layout --- simply place a single vertex on each track. Thus \thmref{DrawingIFFTracknumber} matches the \Oh{n^3} volume bound discussed in \secref{IntroDrawings}. In fact, the drawings of $K_n$ produced by our algorithm, with each vertex in a distinct track, are identical to those produced by the algorithm of Cohen\etal\cite{CELR-Algo96}. 

\twothmref{QueuenumberIFFTracknumber}{DrawingIFFTracknumber} immediately imply the following result, which  reduces the problem of producing a three-dimensional drawing with small volume to that of producing a queue layout of the same graph with few queues.

\begin{theorem}
\thmlabel{DrawingIFFQueue}
Let \G\ be a proper minor-closed family of graphs, and let $\F(n)$ be a family of functions closed under multiplication. The following are equivalent:
\begin{remunerate}
\item[\textup{(a)}] every $n$-vertex graph in \G\ has a $\F(n)\times\F(n)\times \Oh{n}$ drawing,
\item[\textup{(b)}] \G\ has track-number $\tn{\G}\in\F(n)$, and
\item[\textup{(c)}] \G\ has queue-number $\qn{\G}\in\F(n)$.\hfill\endproof
\end{remunerate}
\end{theorem}

Graphs with constant queue-number include  de Bruijn graphs, FFT and Bene\v{s} network graphs \cite{HR-SJC92}. By \thmref{DrawingIFFQueue}, these graphs have three-dimensional drawings with \Oh{n} volume.  Applying \twothmref{TrackLayoutSummary}{DrawingIFFTracknumber} we have the following
result.

\begin{theorem}
\thmlabel{DrawingsSummary}
Let $G$ be a graph with maximum degree $\Delta(G)$,  path-width \pw{G}, tree-partition-width \tpw{G}, and tree-width \tw{G}.  Then $G$ has a three-dimensional drawing with the following dimensions: 
\begin{remunerate}
\item[\textup{(a)}] $\Oh{\pw{G}}\times\Oh{\pw{G}}\times\Oh{n}$, which is 
$\Oh{\tw{G}\,\log n}\times\Oh{\tw{G}\,\log n}\times\Oh{n}$,
\item[\textup{(b)}] $\Oh{\tpw{G}}\times\Oh{\tpw{G}}\times\Oh{n}$, which is
$\Oh{\Delta(G)\,\tw{G}}\times\Oh{\Delta(G)\,\tw{G}}\times\Oh{n}$,
\item[\textup{(c)}] $\Oh{3^{\,\tw{G}}\cdot6^{(4^{\,\tw{G}}-3\,\tw{G}-1)/9}}
\times\Oh{3^{\,\tw{G}}\cdot6^{(4^{\,\tw{G}}-3\,\tw{G}-1)/9}} \times \Oh{n}$.\hfill\endproof
\end{remunerate}
\end{theorem}

Most importantly, we have the following corollary of \thmref{DrawingsSummary}(c).

\begin{corollary}
\corlabel{TreewidthDrawings}
Every graph with bounded tree-width has a three-dimensional drawing with \Oh{n} volume.\hfill\endproof
\end{corollary}

Note that bounded tree-width is not necessary for a graph to have a three-dimensional drawing with \Oh{n} volume. The $\sqrt{n}\times\sqrt{n}$ plane grid graph has $\Theta(\sqrt{n})$ tree-width, and has a $\sqrt{n}\times\sqrt{n}\times 1$ drawing with $n$ volume. It also has a $3$-track layout, and thus, by \lemref{TrackLayout2Drawing}, has a $\Oh{1}\times\Oh{1}\times\Oh{n}$ drawing.

Since a planar graph is $4$-colourable, by the results of Calamoneri and Sterbini~\cite{CS-IPL97} and Pach~\cite{PTT99} discussed above, every planar graph has a  three-dimensional drawing with \Oh{n^2} volume. This result also follows from the classical algorithms of de~Fraysseix\etal\cite{dFPP90} and Schnyder~\cite{Schnyder-Order89} for producing $\Oh{n}\times\Oh{n}$ plane grid drawings. All of these methods produce $\Oh{n}\times\Oh{n}\times\Oh{1}$ drawings, which have $\Theta(n)$ aspect ratio. Since every planar graph $G$ has $\pw{G}\in\Oh{\sqrt{n}}$ \cite{Bodlaender-TCS98}, we have the following corollary of \thmref{DrawingsSummary}(a).

\begin{corollary}
Every planar graph has a three-dimensional drawing with \Oh{n^2} volume and $\Theta(\sqrt{n})$ aspect ratio.\hfill\endproof
\end{corollary}

This result matches the above \Oh{n^2} volume bounds with an improvement in the aspect ratio by a factor  of $\Theta(\sqrt{n})$.  As discussed in \secref{Introduction}, it is an open problem whether every planar graph has a three-dimensional drawing with \Oh{n} volume. Subsequent to this research, Dujmovi{\'c} and Wood~\cite{DujWoo-SubQuad-AMS} proved that graphs excluding a clique minor on a fixed number of vertices, such as planar graphs, have three-dimensional drawings with $\Oh{n^{3/2}}$ volume, as do graphs with bounded degree. 

Our final result regarding three-dimensional drawings, which is proved in \secref{Drawings}, examines the apparent trade-off between aspect ratio and volume.

\begin{theorem}
\thmlabel{AspectRatioDrawings}
For every graph $G$ and for every $r$, $1\leq r \leq n/\tn{G}$, $G$ has a three-dimensional drawing with \Oh{n^3/r^2} volume and aspect ratio $2r$.
\end{theorem}

%%%%%%%%%%%%%%%%%%%%%%%%%%%%%%%%%%%%%%%%%%%%%%%%%%%%%%%%%%%%%%%%%%%%%%%%%%%%%%
\section{Track Layouts}\seclabel{BasicTrackLayouts}
%%%%%%%%%%%%%%%%%%%%%%%%%%%%%%%%%%%%%%%%%%%%%%%%%%%%%%%%%%%%%%%%%%%%%%%%%%%%%%

In this section we describe a number of methods for producing and manipulating track layouts. The following result is implicit in the proof by Felsner\etal\cite{FLW-JGAA03} that every outerplanar graph has an improper $3$-track layout.

\begin{lemma}[\cite{FLW-JGAA03}]
\lemlabel{TreeTrackLayout}
Every tree $T$ has a $3$-track layout.
\end{lemma}

\begin{proof} Root $T$ at an arbitrary node $r$. Let $\sigma$ be a lexicographical breadth-first vertex-ordering of $T$ starting at $r$, as described in \secref{IntroVertexOrderings}. For $i\in\{0,1,2\}$, let $V_i$ be the set of nodes of $T$ with depth $d\equiv i\pmod{3}$ in $\sigma$. With each $V_i$ ordered by $\sigma$, we have a $3$-track assignment of $T$. Clearly adjacent vertices are on distinct tracks. Since no two edges are nested in $\sigma$, there is no X-crossing (see \figref{TreeTrackLayout}).  \hfill\end{proof}

\Figure{TreeTrackLayout}{\includegraphics{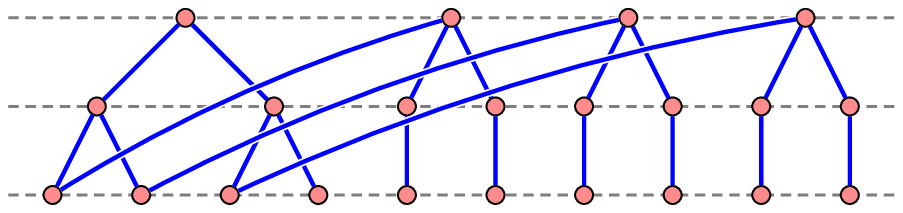}}{A $3$-track
layout of a tree.}

\begin{lemma}
\lemlabel{PathDecomp2TrackLayout}
Every graph $G$ with path-width \pw{G} has track-number $\tn{G}\leq\pw{G}+1$.
\end{lemma}

\begin{proof}
Let $k=\pw{G}+1$. It is well known that a $G$ is the subgraph of a $k$-colourable interval graph \cite{Bodlaender-TCS98,Golumbic80}. That is, there is a set of intervals $\{[\ell(v),r(v)]\subseteq\mathbb{R}:v\in V(G)\}$ such that $[\ell(v),r(v)]\cap[\ell(w),r(w)]\ne\emptyset$ for every edge $vw$ of $G$.  Let $\{V_i:1\leq i\leq k\}$ be a $k$-colouring of $G$.  Consider each colour class $V_i$ to be an ordered track $(v_1,v_2,\dots,v_p)$, where $\ell(v_1)<r(v_1)<\ell(v_2)<r(v_2)<\dots<\ell(v_p)<r(v_p)$. Suppose there is an X-crossing between edges $vw$ and $xy$ with $v,x\in V_i$ and $w,y\in V_j$ for some pair of tracks $V_i$ and $V_j$. Without loss of generality, $r(v)<\ell(x)$ and $r(y)<\ell(w)$.  Since $vw$ is an edge, $\ell(w)\leq r(v)$. Thus $r(y)<\ell(w)\leq r(v)<\ell(x)$, which implies that $xy$ is not an edge of $G$. This contradiction proves that there is no X-crossing, and $G$ has a $k$-track layout.\hfill\end{proof}

\Figure{PathTrackLayout}{\includegraphics{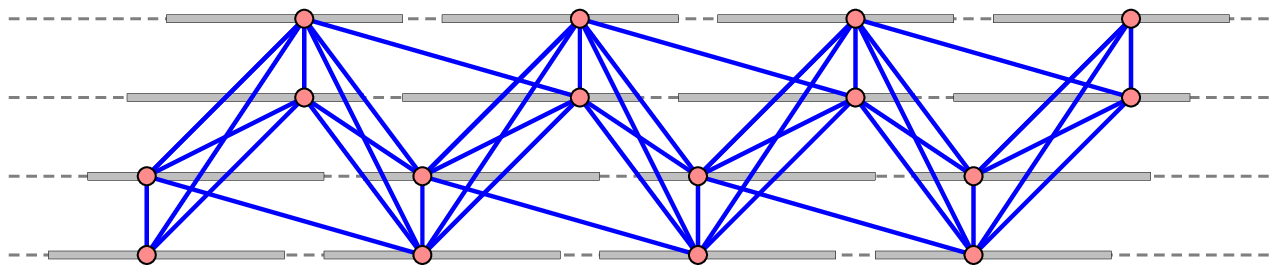}}
{A $4$-track layout of a $4$-colourable interval graph.}

The next lemma uses a tree-partition to construct a track layout.

\begin{lemma}
\lemlabel{TreePartition2TrackLayout}
Every graph $G$ with maximum degree $\Delta(G)$, tree-width \tw{G}, and tree-partition-width \tpw{G}, has track-number $\tn{G}\leq3\,\tpw{G}\leq72\,\Delta(G)\tw{G}$.
\end{lemma} 

\begin{proof}  Let $(T,\{T_x:x \in V(T)\})$ be a tree-partition of $G$ with
width $\tpw{G}$. By \lemref{TreeTrackLayout}, $T$ has a $3$-track layout.
Replace each track by $\tpw{G}$ `sub-tracks', and for each node $x$ in $T$,
place the vertices in bag $T_x$ on the sub-tracks replacing the track
containing $x$, with at most one vertex in $T_x$ in a single track. For all nodes $x$ and $y$ of $T$, if $x<y$ in a single track of the $3$-track layout of $T$, then for all vertices $v\in T_x$ and $w\in T_y$, $v<w$ whenever  $v$ and $w$ are assigned to the same track. There is no X-crossing, since in the track layout of $T$, adjacent nodes are on distinct tracks and there is no X-crossing. Thus  we have a track layout of $G$. The number of tracks is $3\,\tpw{G}$, which is at most  $72\,\Delta(G)\tw{G}$ by the theorem of Ding and Oporowski~\cite{DO-JGT95} discussed in \secref{IntroTreePartition}. \hfill\end{proof}

In the remainder of this section, we prove two results that show how track
layouts can be manipulated without introducing an X-crossing. The first is a
generalisation of the `wrapping' algorithm of Felsner\etal\cite{FLW-JGAA03}, who implicitly proved the case $s=1$.

\begin{lemma}
\lemlabel{Rolling}
If a graph $G$ has an \paran{improper} track layout  $\{(V_i,<_i):1\leq i\leq
t\}$  with maximum edge span $s$, then $G$ has an \paran{improper}
$(2s+1)$-track layout.
\end{lemma}

\begin{proof} Let $\ell=2s+1$. Construct an $\ell$-track assignment of $G$ by merging the tracks $\{V_i:i\equiv j\pmod{t}\}$ for each $j$, $0\leq j\leq t-1$, with vertices in $V_\alpha$ appearing before vertices in $V_\beta$ in the new track $j$, for all $\alpha,\beta\equiv j\pmod{t}$ with $\alpha<\beta$. The given order of each $V_i$ is preserved in the new tracks. It remains to prove that there is no X-crossing. Consider two edges $vw$ and $xy$. Let $i_1$ and $i_2$, $1\leq i_1<i_2\leq t$, be the minimum and maximum tracks  containing $v$, $w$, $x$ or $y$ in the given $t$-track layout of $G$.

First consider the case that $i_2-i_1>2s$. Then without loss of generality
$v$ is in track $i_2$ and $y$ is in track $i_1$. Thus $w$ is in a greater track
than $x$, and even if $x$ (or $y$) appear on the same track as $v$ (or $w$) in
the new $\ell$-track assignment, $x$ (or $y$) will be to the left of $v$ (or $w$). Thus these edges do not form an X-crossing in the  $\ell$-track assignment. 
Otherwise $i_2-i_1\leq2s$. Thus any two of $v$, $w$, $x$ or $y$ will appear on the same track in the $\ell$-track assignment if and only if they are on the same track in the given $t$-track layout (since $\ell>2s$). Hence the only way for these four vertices to appear on exactly two tracks in the  $\ell$-track assignment is if they were on exactly two layers in the given $t$-track layout, in which case, by assumption $vw$ and $xy$ do not form an X-crossing.  Therefore there is no X-crossing, and we have an $\ell$-track layout of $G$.
\hfill\end{proof}

The next result shows that the number of vertices in different tracks of a
track layout can be balanced without introducing an X-crossing. The proof is
based on an idea due to Pach\etal\cite{PTT99} for balancing the size of the
colour classes in a colouring.

\begin{lemma}
\lemlabel{GeneralBalancing}
If a graph $G$ has an \paran{improper}  $t$-track layout, then for every
$t'>0$, $G$  has an \paran{improper} $\floor{t+t'}$-track layout with at
most $\ceil{\frac{n}{t'}}$ vertices in each track.
\end{lemma}

\begin{proof}   For each track with $q>\ceil{\frac{n}{t'}}$ vertices,
replace it by $\ceil{q/\ceil{\frac{n}{t'}}}$ `sub-tracks' each with exactly
$\ceil{\frac{n}{t'}}$ vertices except for at most one sub-track with
$q\bmod{\ceil{\frac{n}{t'}}}$ vertices, such that the vertices in each
sub-track are consecutive in the original track, and the original order is
maintained. There is no X-crossing between sub-tracks from the same original
track as there is at most one edge between such sub-tracks. There is no
X-crossing between sub-tracks from different original tracks as otherwise there
would be an X-crossing in the original. There are at most $\floor{t'}$
tracks with $\ceil{\frac{n}{t'}}$ vertices. Since there are at most $t$
tracks with less than $\ceil{\frac{n}{t'}}$ vertices, one for each of the
original tracks, there is a total of at most $\floor{t+t'}$ tracks. 
\hfill\end{proof}

%%%%%%%%%%%%%%%%%%%%%%%%%%%%%%%%%%%%%%%%%%%%%%%%%%%%%%%%%%%%%%%%%%%%%%%%%%%%%%%
\section{Three-Dimensional Drawings and Track Layouts}
\seclabel{Drawings}
%%%%%%%%%%%%%%%%%%%%%%%%%%%%%%%%%%%%%%%%%%%%%%%%%%%%%%%%%%%%%%%%%%%%%%%%%%%%%%%

In this section we prove \thmref{DrawingIFFTracknumber}, which states that
three-dimensional drawings with small volume are closely related to track
layouts with few tracks.

\begin{lemma}
\lemlabel{Drawing2TrackLayout}
If a graph $G$ has an $A\times B\times C$  drawing, then $G$ has an improper
$AB$-track layout, and $G$ has a $2AB$-track layout.
\end{lemma}

\begin{proof} Let $V_{x,y}$ be the set of vertices of $G$ with an
$X$-coordinate of $x$ and a $Y$-coordinate of $y$, where without loss of
generality $1\leq x\leq A$ and $1\leq y\leq Y$. With each set $V_{x,y}$ 
ordered by the $Z$-coordinates of its elements, $\{V_{x,y}:1\leq x\leq
A,1\leq y\leq Y\}$ is an improper $AB$-track assignment. There is no X-crossing,
as otherwise there would be a crossing in the original drawing, and hence we
have an improper $AB$-track layout.  By \lemref{ImproperProper}, $G$ has a
$2AB$-track layout. \hfill\end{proof}

We now prove the converse of \lemref{Drawing2TrackLayout}.  The proof is
inspired by the generalisations of the moment curve algorithm by
Cohen\etal\cite{CELR-Algo96} and Pach\etal\cite{PTT99}, described in
\secref{IntroDrawings}.  Loosely speaking, Cohen\etal\cite{CELR-Algo96} allow three `free' dimensions, whereas Pach\etal\cite{PTT99} use the assignment of vertices to colour classes to `fix' one dimension with two dimensions free. We use an assignment of vertices to tracks to fix two dimensions with one dimension free.  The style of three-dimensional drawing produced by our algorithm, where tracks are drawn vertically, is illustrated in \figref{ThreeDimDrawing}.

\Figure{ThreeDimDrawing}{\includegraphics{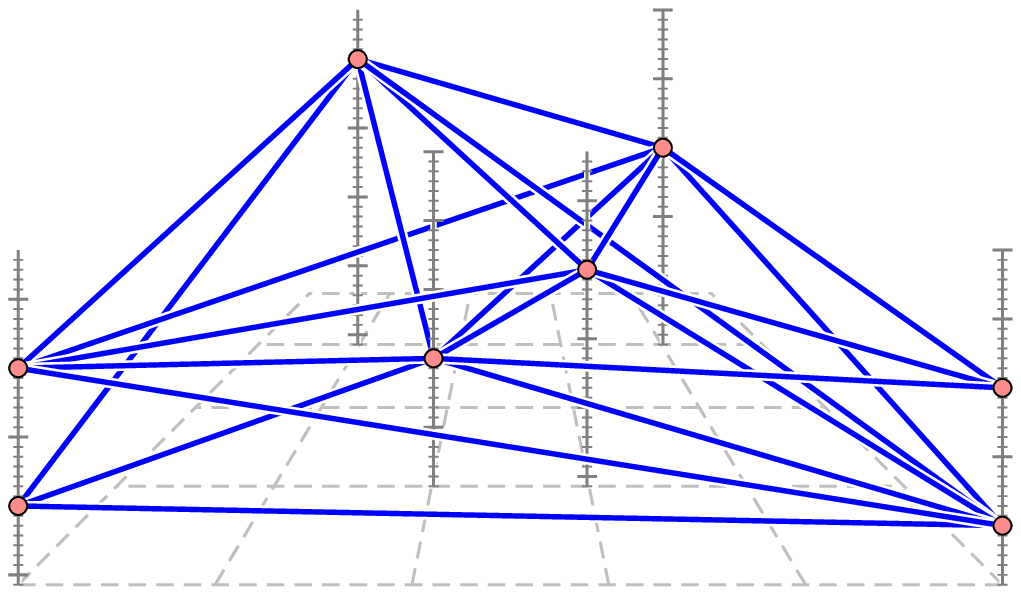}}{A
three-dimensional drawing produced from a track layout.}

\begin{lemma}
\lemlabel{TrackLayout2Drawing}
If a graph $G$ has a \paran{possibly} improper $k$-track layout, then $G$ has a
$k\times 2k\times 2k\cdot n'$ three-dimensional drawing, where $n'$ is the
maximum number of vertices in a track.
\end{lemma}

\begin{proof} 
Suppose $\{(V_i,<_i):1\leq i\leq k\}$ is the given improper $k$-track layout.
Let $p$ be the smallest prime such that $p> k$. Then $p\leq2k$ by Bertrand's
postulate. For each $i$, $1\leq i\leq k$, represent the vertices in $V_i$ by
the grid-points
\begin{equation*}
\{(i,i^2\bmod{p},t):\,1\leq t\leq p\cdot|V_i|,\,t\equiv i^3\!\!\pmod{p}\}\enspace,
\end{equation*}
such that the $Z$-coordinates respect the given total order $<_i$. Draw
each edge as a line-segment between its end-vertices.  Suppose two
edges $e$ and $e'$ cross such that their end-vertices are at distinct points
$(i_\alpha,i_\alpha^2\bmod{p},t_\alpha)$, $1\leq \alpha\leq 4$. 
Then these points are coplanar, and if $M$ is the matrix
\begin{equation*}
M=
\begin{pmatrix}
1	&\;\; i_1	&\;\; i_1^2\bmod{p}	&\;\; t_1 \\
1	&\;\; i_2	&\;\; i_2^2\bmod{p}	&\;\; t_2 \\
1	&\;\; i_3	&\;\; i_3^2\bmod{p}	&\;\; t_3 \\
1	&\;\; i_4	&\;\; i_4^2\bmod{p}	&\;\; t_4 \\
\end{pmatrix}
\end{equation*}
then the determinant $\text{det}(M)=0$. We proceed by considering the number of
distinct tracks $N=|\{i_1,i_2,i_3,i_4\}|$. 

$\bullet$ $N=1$: By the definition of an improper track layout, 
$e$ and $e'$ do not cross. 

$\bullet$ $N=2$: If either edge is intra-track   then $e$ and $e'$ do not
cross. Otherwise neither edge is intra-track,  and since there is no
X-crossing, $e$ and $e'$ do not cross. 

$\bullet$ $N=3$: Without loss of generality $i_1=i_2$. It follows that
$\text{det}(M) = (t_2-t_1)\cdot \text{det}(M')$, where 
\begin{equation*}
M'=
\begin{pmatrix}
1 &\;\; i_2 &\;\; i_2^2 \bmod{p}	\\
1 &\;\; i_3 &\;\; i_3^2 \bmod{p} 	\\
1 &\;\; i_4 &\;\; i_4^2 \bmod{p}	\\
\end{pmatrix}
\enspace.
\end{equation*}
Since $t_1\ne t_2$, $\text{det}(M')=0$. 
However, $M'$ is a Vandermonde matrix modulo $p$, and
thus
\begin{equation*}
\text{det}(M')\,\equiv\,(i_2-i_3)(i_2-i_4)(i_3-i_4)\;\pmod{p},
\end{equation*}
which is non-zero since $i_2$, $i_3$ and $i_4$ are distinct and $p$ is a prime,
a  contradiction.

$\bullet$ $N=4$: Let $M'$ be the matrix obtained from
$M$ by taking each entry modulo $p$. Then $\text{det}(M')=0$.  
Since $t_\alpha\equiv i_\alpha^3\pmod{p}$, $1\leq \alpha\leq 4$, 
\begin{equation*}
M'\equiv
\begin{pmatrix}
1	&\;\; i_1	&\;\; i_1^2	&\;\; i_1^3 \\
1	&\;\; i_2	&\;\; i_2^2	&\;\; i_2^3 \\
1	&\;\; i_3	&\;\; i_3^2	&\;\; i_3^3 \\
1	&\;\; i_4	&\;\; i_4^2	&\;\; i_4^3 \\
\end{pmatrix}
\pmod{p}\enspace.
\end{equation*}
Since each $i_\alpha<p$, $M'$ is a Vandermonde matrix modulo $p$, and thus
\begin{equation*}
\text{det}(M')\,\equiv\,(i_1-i_2)(i_1-i_3)(i_1-i_4)(i_2-i_3)(i_2-i_4)(i_3-i_4)\;\pmod{p},
\end{equation*}
which is non-zero since $i_\alpha\ne i_\beta$ and $p$ is a prime. This 
contradiction proves there are no edge crossings. The produced 
drawing is at most $k\times 2k\times 2k\cdot n'$.
\hfill\end{proof}

\emph{Proof of \thmref{DrawingIFFTracknumber}}. Let $\F(n)$ be
a family of functions closed under multiplication. Let $G$ be an
$n$-vertex graph with a $t$-track layout, where $t\in\F(n)$. By
\lemref{GeneralBalancing} with $t'=t$, $G$ has a $2t$-track layout with at most
$\ceil{\frac{n}{t}}$ vertices in each track. By \lemref{TrackLayout2Drawing},
$G$ has a $2t\times 4t\times 4t\cdot\ceil{\frac{n}{t}}$ drawing, which is
$\Oh{t}\times\Oh{t}\times\Oh{n}$. Conversely, suppose an $n$-vertex graph $G$ has a $A\times B\times \Oh{n}$ drawing, where $A,B\in\F(n)$. By
\lemref{Drawing2TrackLayout}, $G$ has a track layout with
$2AB\in\F(n)$ tracks. \hfill\endproof

\emph{Proof of \thmref{AspectRatioDrawings}}. Let $t=\tn{G}$, and
suppose $1\leq r \leq n/t$. By \lemref{GeneralBalancing} with $t'=\frac{n}{r}$,
$G$  has a $\floor{\frac{n}{r}+t}$-track layout with at most $r$ vertices in
each track. By assumption $t\leq\frac{n}{r}$, and the number of tracks is at
most $\frac{2n}{r}$. By \lemref{TrackLayout2Drawing}, $G$ has a $\frac{2n}{r}
\times \frac{4n}{r} \times 4n$ three-dimensional drawing, which has volume
$32n^3/r^2$ and aspect ratio $2r$.  \hfill\endproof

%%%%%%%%%%%%%%%%%%%%%%%%%%%%%%%%%%%%%%%%%%%%%%%%%%%%%%%%%%%%%%%%%%%%%%%%%%%
\section{Queue Layouts and Track Layouts}
\seclabel{QueueLayouts}
%%%%%%%%%%%%%%%%%%%%%%%%%%%%%%%%%%%%%%%%%%%%%%%%%%%%%%%%%%%%%%%%%%%%%%%%%%%%%%%

In this section we prove \thmref{QueuenumberIFFTracknumber}, which states that
track and queue layouts are closely related. Our first lemma highlights this
fact --- its proof follows immediately from the definitions (see
\figref{BipartiteLayout}).

\begin{lemma}
\lemlabel{BipartiteLayout}
A bipartite graph $G=(A,B;E)$ has a $2$-track layout with tracks $A$ and $B$ 
if and only if $G$ has a $1$-queue layout such that in the corresponding
vertex-ordering, the vertices in $A$ appear before the vertices in $B$.\hfill\endproof
\end{lemma}

\Figure{BipartiteLayout}{\includegraphics{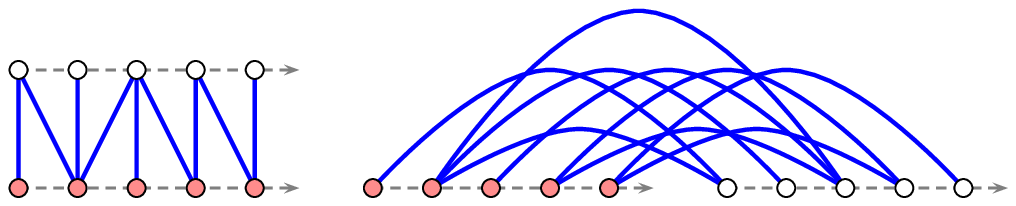}}
{A $2$-track layout and a $1$-queue layout of a bipartite graph.}

We now show that a queue layout can be obtained from a track layout. This
result can be viewed as a generalisation of the construction of a $2$-queue
layout of an outerplanar graph by Heath\etal\cite{HLR-SJDM92} and Rengarajan and Veni~Madhavan~\cite{RM-COCOON95} (with $s=1$).

\begin{lemma} 
\lemlabel{TrackLayout2QueueLayout} 
If a graph $G$ has a \paran{possibly} improper $t$-track layout
$\{(V_i,<_i):1\leq i\leq t\}$ with maximum edge span $s$ \paran{$\leq t-1$}, then $\qn{G}\leq s+1$, and if the given track layout is not improper, then $\qn{G}\leq s$. 
\end{lemma}

\begin{proof} First suppose that there are no intra-track edges. Let $\sigma$ be the vertex ordering $(V_1,V_2,\dots,V_t)$ of $G$. Let $E_\alpha$ be the set of edges with span $\alpha$ in the given track layout. As in \lemref{BipartiteLayout}, two edges from the same pair of tracks are nested in $\sigma$ if and only if they form an X-crossing in the track layout. Since no two edges form an X-crossing in the track layout, no two edges that are between the same pair of tracks are nested in $\sigma$. If two edges not from the same pair of tracks have the same span then they are not nested in $\sigma$. (This idea is due to Heath and Rosenberg~\cite{HR-SJC92}.)\ Thus no two edges are nested in each $E_\alpha$, and we have an $s$-queue layout of $G$. If there are intra-track edges, then they all form one additional queue in $\sigma$.
\hfill\end{proof}

We now set out to prove the converse of \lemref{TrackLayout2QueueLayout}. 
It is well known that the subgraph induced by any two tracks  of a track layout is a forest of caterpillars \cite{HS72}. A colouring of a graph is \emph{acyclic} if every bichromatic subgraph is a forest; that is, every cycle receives at least three distinct colours. Thus a $t$-track layout of a graph $G$ defines an acyclic $t$-colouring of $G$.  The minimum number of colours in an acyclic colouring of $G$ is the \emph{acyclic chromatic number} of $G$, denoted by \acn{G}. Thus,
\begin{equation*}
\acn{G}\;\leq\;\tn{G}\enspace.
\end{equation*}

Acyclic colourings were introduced by Gr{\"u}nbaum~\cite{Grunbaum73}, who proved that every planar graph is acyclically $9$-colourable. This result was steadily improved \cite{AB-IJM77, Kostochka76, Mitchem-DMJ74} until Borodin~\cite{Borodin-DM79} proved that every planar graph is acyclically $5$-colourable, which is the best possible bound. Many other graph families have bounded acyclic chromatic number, including graphs embeddable on a fixed surface \cite{AB-GMJ78, Albertson-EJC04, AMS-IJM96}, $1$-planar graphs  \cite{BKRS01}, graphs with bounded maximum degree \cite{AMR-RSA91}, and graphs with bounded tree-width. A folklore result states that $\acn{G}\leq\tw{G}+1$ (see \cite{FRR-WG01}). More generally, Ne\v{s}et\v{r}il and Ossona de Mendez~\cite{NesOdM-03} proved that  every proper minor-closed graph family has bounded acyclic chromatic number. In fact, Ne\v{s}et\v{r}il and Ossona de Mendez~\cite{NesOdM-03} proved that every graph $G$ has a \emph{star} $k$-colouring (every bichromatic subgraph is a forest of stars), where $k$ is a (small) quadratic function of the maximum chromatic number of a minor of $G$. 

\begin{lemma}
\lemlabel{QueueLayout2TrackLayout}
Every graph $G$ with acyclic chromatic number $\acn{G}\leq c$ and queue-number $\qn{G}\leq q$ has track-number $\tn{G}\leq c\,(2q)^{c-1}$. 
\end{lemma}

\begin{proof}  
Let $\{V_i:1\leq i\leq c\}$ be an acyclic colouring of $G$. Let $\sigma$ be the vertex-ordering in a $q$-queue layout of $G$. Consider an edge $vw$ with $v\in V_i$, $w\in V_j$, and $i<j$. If $v<_\sigma w$ then $vw$ is \emph{forward}, and if $w<_\sigma v$ then $vw$ is \emph{backward}. Consider the edges to be coloured with $2q$ colours, where each colour class consists of the forward edges in a single queue, or the backward edges in a single queue.

Alon and Marshall~\cite{AM-JAC98} proved that given a (not necessarily  proper) edge $k$-colouring of a graph $G$, any acyclic $c$-colouring of $G$ can be refined to a $ck^{c-1}$-colouring so that  the edges between  any pair of (vertex) colour classes are monochromatic, and each (vertex) colour class is contained in some original colour class. (Ne\v{s}et\v{r}il and Raspaud~\cite{NR-JCTB00} generalised this result for coloured mixed graphs.) Apply this result with the given acyclic $c$-colouring of $G$ and the edge  $2q$-colouring discussed above. Consider the resulting $c(2q)^{c-1}$  colour classes to be tracks ordered by $\sigma$. The edges between any two  tracks are from a single queue, and are all forward or all backward. 

Suppose that there are edges $vw$ and $xy$ that form an X-crossing. Since each track is a subset of some $V_i$, we can  assume that $v,x\in V_i$, $w,y\in V_j$ and $i<j$. Suppose that $vw$ and  $xy$ are both forward. The case in which $vw$ and $xy$ are both backward is symmetric. Thus $v<_\sigma w$ and  $x<_\sigma y$.  Since $vw$ and $xy$  form an X-crossing, and the tracks are ordered by $\sigma$, we have $v<_\sigma x$ and $y<_\sigma w$. Hence $v<_\sigma x<_\sigma y<_\sigma w$. That is, $vw$ and $xy$ are nested. This is the desired contradiction, since edges between any pair of tracks are from a single queue. Thus we have a $c(2q)^{c-1}$-track layout of $G$.
\hfill\end{proof}

\emph{Proof of \thmref{QueuenumberIFFTracknumber}}. Let $\F(n)$ be a family of functions closed under multiplication. Let $G$ be an $n$-vertex graph from a proper minor-closed graph family \G.  First, suppose that $G$ has a $t$-track layout, where $t\in\F(n)$. By \lemref{TrackLayout2QueueLayout}, $G$ has queue-number $\qn{G}\leq t-1\in \F(n)$. Conversely, suppose $G$ has queue-number $\qn{G}=q\in\F(n)$. By the above-mentioned result of Ne{\v{s}}et{\v{r}}il and Ossona de Mendez~\cite{NesOdM-03}, $G$ has bounded acyclic chromatic number $\acn{G}\leq c\in \Oh{1}$.  By \lemref{QueueLayout2TrackLayout},  $G$ has a $t$-track layout, where $t\,\leq\,c(2q)^{c-1}\in\,\F(n)$.  \hfill\endproof

%%%%%%%%%%%%%%%%%%%%%%%%%%%%%%%%%%%%%%%%%%%%%%%%%%%%%%%%%%%%%%%%%%%%%%%%%%%%%%%%
\section{\boldmath Tree-Partitions of $k$-Trees}
\seclabel{TreePartitions}
%%%%%%%%%%%%%%%%%%%%%%%%%%%%%%%%%%%%%%%%%%%%%%%%%%%%%%%%%%%%%%%%%%%%%%%%%%%%%%%%

In this section we prove our theorem regarding tree-partitions of $k$-trees
mentioned in \secref{IntroTreePartition}. This result forms the cornerstone of
the proof of \thmref{TreewidthTracknumber}.

\begin{theorem}
\thmlabel{TreePartition}
Let $G$ be a $k$-tree with maximum degree $\Delta$. 
Then $G$ has a rooted tree-partition $(T,\{T_x:x\in V(T)\})$ such
that for all nodes $x$ of $T$,
\begin{remunerate}
\item[\textup{(a)}] if $x$ is a non-root node of $T$ and $y$ is the parent
node of $x$,  then the set of vertices in $T_y$ with a neighbour in $T_x$ form
a clique $C_x$ of $G$, and
\item[\textup{(b)}] the induced subgraph $G[T_x]$ is a connected
$(k-1)$-tree.
\end{remunerate}
Furthermore the width of $(T,\{T_x:x\in V(T)\})$ is at most
$\max\{1,k(\Delta-1)\}$. 
\end{theorem}

\begin{proof} We assume $G$ is connected, since  if $G$ is not connected then a
tree-partition of $G$ that satisfies the theorem can be determined by adding a
new root node with an empty bag, adjacent to the root node of a tree-partition of each connected component of $G$. 

It is well-known that $G$ is a connected $k$-tree if and only if $G$ has a vertex-ordering $\sigma=(v_1,v_2,\dots,v_n)$, such that for all $i\in\{1,2,\dots,n\}$,
\begin{romannum}
\item if $G^i$ is the induced subgraph
$G[\{v_1,v_2,\dots,v_i\}]$, then $G^i$ is connected and the vertex-ordering of
$G^i$ induced by $\sigma$ is a breadth-first vertex-ordering of $G^i$, and
\item the neighbours of $v_i$ in $G^i$ form a clique
$C_i=\{v_j:v_iv_j\in E(G),j<i\}$ with  $1\leq|C_i|\leq k$ (unless $i=1$ in
which case $C_i=\emptyset$).
\end{romannum}

In the language of chordal graphs, $\sigma$ is a (reverse) `perfect
elimination' vertex-ordering and can be determined,  for example, by the
Lex-BFS algorithm by Rose\etal\cite{RTL-SJC76} (also see \cite{Golumbic80}).
Moreover, we can choose $v_1$ to be any vertex in $G$.

Let $r$ be a vertex of minimum degree\footnote{We choose $r$ to have minimum
degree to obtain a slightly improved bound on the width of the
tree-partition. If we choose $r$ to be an arbitrary vertex then the width is at
most  $\max\{1,\Delta,k(\Delta-1)\}$, and the remainder of
\thmref{TreePartition} holds.} in $G$. Then $\deg(r)\leq k$. Let
$\sigma=(v_1,v_2,\dots,v_n)$ be a vertex-ordering of $G$ with $v_1=r$, and
satisfying (i) and (ii). By (i), the depth of each vertex $v_i$ in $\sigma$ is
the same as the depth of $v_i$ in the vertex-ordering of $G^j$ induced by
$\sigma$, for all $j\geq i$.  We therefore simply speak of \emph{the} depth of
$v_i$. Let $V_d$ be the set of vertices of $G$ at depth $d$. 

\begin{claim}
\label{TheClaim}
For all $d\geq1$, and for every connected component  $Z$ of $G[V_d]$, the set of vertices at depth $d-1$ with a neighbour in $Z$ form a clique of $G$.
\end{claim}

\begin{proof}
The claim in trivial for $d=1$ or $d=2$. Now suppose that $d\geq 3$. Assume for  the sake of contradiction that there are two  non-adjacent vertices $x$ and $y$ at depth $d-1$, such that $x$ has a neighbour in $Z$ and $y$ has a neighbour in $Z$. Let $P_1$ be a shortest path between $x$ and $y$ with its interior vertices in $Z$. Let $P_2$ be a shortest path between $x$ and $y$ with its interior vertices at depth at most $d-2$. Since the interior vertices of $P_1$ are at depth $d$, there is no edge between an interior vertex of $P_1$ and an interior vertex of $P_2$. Thus $P_1 \cup P_2$ is a chordless cycle of length at least four, contradicting the fact that $G$ is chordal (by \lemref{TreewidthCharacterisation}).
\hfill\end{proof}

Define a graph $T$ and a partition $\{T_x:x\in V(T)\}$ of $V(G)$ indexed by the
nodes of $T$ as follows.  There is one node $x$ in $T$ for every connected
component of each $G[V_d]$, whose  bag $T_x$ is the vertex-set of the
corresponding connected component.  We say $x$ and $T_x$ are at \emph{depth}
$d$.  Clearly a vertex in a depth-$d$ bag is also at depth $d$. The (unique)
node of $T$ at depth zero is called the \emph{root} node. Let two nodes $x$ and
$y$ of $T$ be connected by an edge if there is an edge $vw$ of $G$ with $v\in
T_x$ and $w\in T_y$. Thus $(T,\{T_x:x\in V(T)\})$ is a `graph-partition'.

We now prove that in fact $T$ is a tree. First observe that $T$ is connected
since $G$ is connected. By definition, nodes of $T$ at the same depth $d$ are
not adjacent. Moreover nodes of $T$ can be adjacent only if their depths differ
by one. Thus $T$ has a cycle only if there is a node $x$ in $T$ at some depth
$d$, such that $x$ has at least two distinct neighbours in $T$ at depth $d-1$.
However this is impossible since by  Claim~\ref{TheClaim}, the set of vertices at depth $d-1$ with a neighbour in $T_x$ form a clique (which we call $C_x$), and are hence in a single bag at depth $d-1$. Thus $T$ is a tree and $(T,\{T_x:x\in V(T)\})$ is a tree-partition of $G$ (see \figref{TreePartition}).

\Figure{TreePartition}{\includegraphics{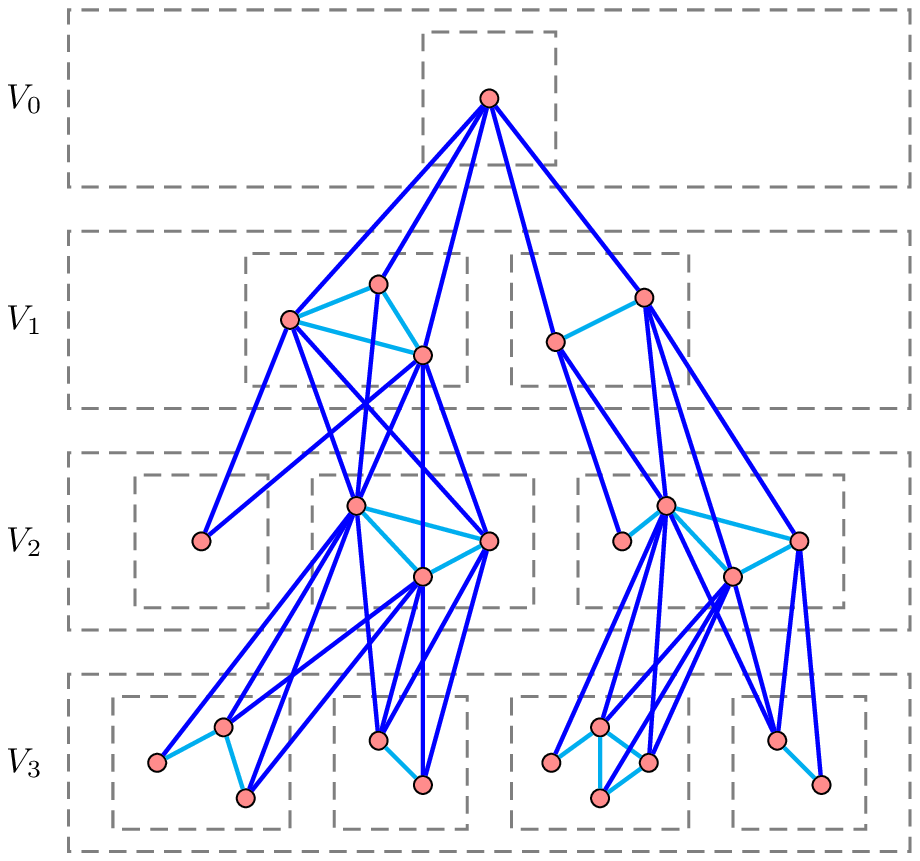}}{Illustration for
\thmref{TreePartition} in the case of $k=3$.}

We now prove that each bag $T_x$ induces a connected $(k-1)$-tree. This is true
for the root node which only has one vertex. Suppose $x$ is a non-root node of
$T$ at depth $d$. Each vertex in $T_x$ has at least one neighbour at depth
$d-1$. Thus in the vertex-ordering of $T_x$ induced by $\sigma$, each vertex
$v_i\in T_x$ has at most $k-1$ neighbours $v_j\in T_x$ with $j<i$. 
Thus the vertex-ordering of $T_x$ induced by $\sigma$ satisfies (i) and (ii) for $k-1$, and $G[T_x]$ is $(k-1)$-tree. By definition each $G[T_x]$ is connected.

Finally, consider the cardinality of a bag in $T$. We claim that each bag contains at most  $\max\{1,k(\Delta-1)\}$ vertices. The root bag has one vertex. Let $x$ be a non-root node of $T$ with parent node $y$.  Suppose $y$ is the root node. Then $T_y=\{r\}$, and thus $|T_x|\leq\deg(r)\leq k\leq k(\Delta-1)$ assuming $\Delta\geq2$. If $\Delta\leq1$ then all bags have one vertex. Now assume $y$ is a non-root node.  The set of vertices in $T_y$ with a neighbour in $T_x$ forms the clique $C_x$.  Let $k'=|C_x|$. Thus $k'\geq1$, and since $C_x\subseteq T_y$ and $G[T_y]$ is a $(k-1)$-tree, $k'\leq k$. A vertex $v\in C_x$ has $k'-1$ neighbours in $C_x$ and at least one neighbour in the parent bag of $y$. Thus $v$ has at most $\Delta-k'$ neighbours in $T_x$. Hence the number of edges between $C_x$ and $T_x$ is at most $k'(\Delta-k')$.  Every vertex in $T_x$ is adjacent to a vertex in $C_x$. Thus $|T_x|\leq k'(\Delta-k')\leq k(\Delta-1)$. This completes the proof.\hfill\end{proof}

%%%%%%%%%%%%%%%%%%%%%%%%%%%%%%%%%%%%%%%%%%%%%%%%%%%%%%%%%%%%%%%%%%%%%%%%%%%%%%
\section{Tree-Width and Track Layouts}
\seclabel{TreewidthTrackLayout}
%%%%%%%%%%%%%%%%%%%%%%%%%%%%%%%%%%%%%%%%%%%%%%%%%%%%%%%%%%%%%%%%%%%%%%%%%%%%%%

In this section we prove that track-number is bounded by tree-width. Let $\{(V_i,<_i):i\in I\}$ be a track layout of a graph $G$.  We say a clique $C$ of $G$ \emph{covers} the set of tracks $\{i\in I: C\cap V_i\ne\emptyset\}$.  Let $S$ be a set of cliques of $G$. Suppose there exists a total order $\preceq$ on $S$ such that for all cliques $C_1,C_2\in S$, if there exists a track $i\in I$, and vertices $v\in V_i\cap C_1$ and $w\in V_i \cap C_2$ with $v<_iw$, then $C_1\prec C_2$. In this case, we say $\preceq$ is \emph{nice}, and $S$ is \emph{nicely ordered} by the track layout. 

\begin{lemma}\lemlabel{NiceOrdering}
Let $L\subseteq I$ be a set of tracks in a track layout $\{(V_i,<_i):i\in I\}$ of a graph $G$. If $S$ is a set of cliques, each of which covers $L$, then $S$ is nicely ordered by the given track layout.
\end{lemma}

\begin{proof}  Define a relation $\preceq$ on $S$ as follows. For every pair of  cliques $C_1,C_2\in S$, define $C_1\preceq C_2$ if $C_1=C_2$ or there exists a track $i\in L$ and vertices $v\in C_1$ and $w\in C_2$ with $v<_iw$. Clearly all cliques in $S$ are comparable.

Suppose that $\preceq$ is not antisymmetric; that is, there exists distinct cliques $C_1,C_2\in S$, distinct tracks $i,j\in L$, and distinct vertices $v_1,w_1\in C_1$ and $v_2,w_2\in C_2$, such that $v_1<_iv_2$ and $w_2<_jw_1$. Since $C_1$ and $C_2$ are cliques, the edges $v_1w_1$ and $v_2w_2$ form an X-crossing, which is a contradiction. Thus $\preceq$ is antisymmetric. 

We claim that $\preceq$ is transitive. Suppose there exist cliques $C_1,C_2,C_3\in S$ such that $C_1\preceq C_2$ and $C_2\preceq C_3$.  We can assume that $C_1$, $C_2$ and $C_3$ are pairwise distinct. Thus there are vertices $u_1\in C_1$, $u_2\in C_2$, $v_2\in C_2$ and  $v_3\in C_3$, such that $u_1<_iu_2$ and $v_2<_jv_3$ for some pair of (not necessarily distinct) tracks  $i,j\in L$. Since $C_3$ has a vertex in $V_i$ and since $C_3\not\preceq C_2$, there is a vertex $u_3\in C_3$ with $u_2\leq_i u_3$. Thus $u_1<_iu_3$, which implies that $C_1\preceq C_3$. Thus $\preceq$ is transitive. 

Hence $\preceq$ is a total order on $S$, which by definition is nice.
\hfill\end{proof}

Consider the problem of partitioning the cliques of a graph into sets such that each set is nicely ordered by a given track layout.  The following immediate corollary of \lemref{NiceOrdering} says that there exists such a partition where the number of sets does not depend upon the size of the graph.

\begin{corollary}
\corlabel{CliquePartition}
Let $G$ be a graph with maximum clique size $k$. Given a  $t$-track layout of $G$, there is a partition of the cliques of $G$ into $\sum_{i=1}^k\binom{t}{i}$ sets, each of which is nicely ordered by the given track layout.\hfill\endproof
\end{corollary}

%%%%%%%%%%%%%%%%%%%%%%%%%%%%%%%%%%%%%%%%%%%%%%%%%%%%%%%%%%%%%%%%%%%%%%%%%%%%%

We do not actually use \corref{CliquePartition} in the following result, but the idea of partitioning the cliques into nicely ordered sets  is central to its proof.

\begin{theorem}
\thmlabel{TreewidthTracknumber}
For every integer $k\geq0$, there is a constant $t_k=3^k\cdot6^{(4^k-3k-1)/9}$ such that every graph $G$ with tree-width $\tw{G}\leq k$ has a $t_k$-track layout.
\end{theorem}

\begin{proof}  If the input graph $G$ is not a $k$-tree then add edges to $G$ to obtain a $k$-tree containing $G$ as a subgraph. It is well-known that a graph with tree-width at most $k$ is a \emph{spanning} subgraph of a $k$-tree. These extra edges can be deleted once we are done. We proceed by induction on $k$ with the following hypothesis:

\emph{For all $k\in\mathbb{N}$, there exists a constant $s_k$, and
sets $\III_k$ and $\SSS_k$ such that
\begin{remunerate}
\item $|\III_k|=t_k$ and $|\SSS_k|=s_k$,
\item each element of $\SSS_k$ is a subset of $\III_k$, and
\item every $k$-tree $G$ has a $t_k$-track layout indexed by $\III_k$, such that
for every clique $C$ of $G$, the set of tracks that $C$ covers is in $\SSS_k$.
\end{remunerate}}

Consider the base case with $k=0$. A $0$-tree $G$ has no edges and thus has a $1$-track layout. Let $\III_0=\{1\}$ and order $V_1=V(G)$ arbitrarily. Thus $t_0=1$, $s_0=1$, and $\SSS_0=\{\{1\}\}$ satisfy the hypothesis for every $0$-tree. Now suppose the result holds for $k-1$, and $G$ is a $k$-tree. 

Let $(T,\{T_x:x\in V(T)\})$ be a tree-partition of $G$ described in \thmref{TreePartition}, where $T$ is rooted at $r$. Each induced subgraph $G[T_x]$  is a $(k-1)$-tree. Thus, by induction, there are sets $\III_{k-1}$ and $\SSS_{k-1}$ with $|\III_{k-1}|=t_{k-1}$ and $|\SSS_{k-1}|=s_{k-1}$, such that for every node $x$ of $T$, the induced subgraph $G[T_x]$ has a $t_{k-1}$-track layout indexed by $\III_{k-1}$. For every clique $C$ of $G[T_x]$, if $C$ covers $L\subseteq\III_{k-1}$ then $L\in\SSS_{k-1}$. Assume $\III_{k-1}=\{1,2,\dots,t_{k-1}\}$ and $\SSS_{k-1}=\{X_1,X_2,\dots,X_{s_{k-1}}\}$. By \thmref{TreePartition}, for each non-root node $x$ of $T$, if $p$ is the parent node of $x$, then the set of vertices in $T_p$ with a neighbour in $T_x$ form a clique $C_x$. Let $\alpha(x)=i$ where $C_x$ covers $X_i$. For the root node $r$ of $T$, let $\alpha(r)=1$.

%%%%%%%%%%%%%%%%%%%%%%%%%%%%%%%%%%%%%%%%%%%%%%%%%%
\subsubsection*{Track layout of {\boldmath $T$}}

To construct a track layout of $G$ we first construct a track layout of the tree $T$ indexed by the set $\{(d,i):d\geq0,1\leq i\leq s_{k-1}\}$, where the track $L_{d,i}$ consists of nodes $x$ of $T$ at depth $d$ with $\alpha(x)=i$.  Here the \emph{depth} of a node $x$ is the distance in $T$ from the root node $r$ to $x$. We order the nodes of $T$ within the tracks by increasing depth.  There is only one node at depth $d=0$. Suppose we have determined the orders of the nodes up to depth $d-1$ for some $d\geq1$.

Let $i\in\{1,2,\dots,s_{k-1}\}$. The nodes in $L_{d,i}$ are ordered primarily with respect to the relative positions of their parent nodes (at depth $d-1$). More precisely, let $\rho(x)$ denote the parent node of each node $x\in L_{d,i}$. For all nodes $x$ and $y$ in $L_{d,i}$, if $\rho(x)$ and $\rho(y)$ are in the same track and $\rho(x)<\rho(y)$ in that track, then $x<y$ in $L_{d,i}$. For $x$ and $y$ with $\rho(x)$ and $\rho(y)$ on distinct tracks, the relative order of $x$ and $y$ is not important. It remains to specify the order of nodes in $L_{d,i}$ with a common parent.

Suppose $P$ is a set of nodes in $L_{d,i}$ with a common parent node $p$. By construction, for every node $x\in P$, the parent clique $C_x$ covers $X_i$  in the track layout of $G[T_p]$. By \lemref{NiceOrdering} the cliques $\{C_x:x\in P\}$ are nicely ordered by the track layout of $G[T_p]$. Let the order of $P$ in track $L_{d,i}$ be specified by a nice ordering of $\{C_x:x\in P\}$, as illustrated in \figref{ChildBags}. 

\Figure{ChildBags}{\includegraphics{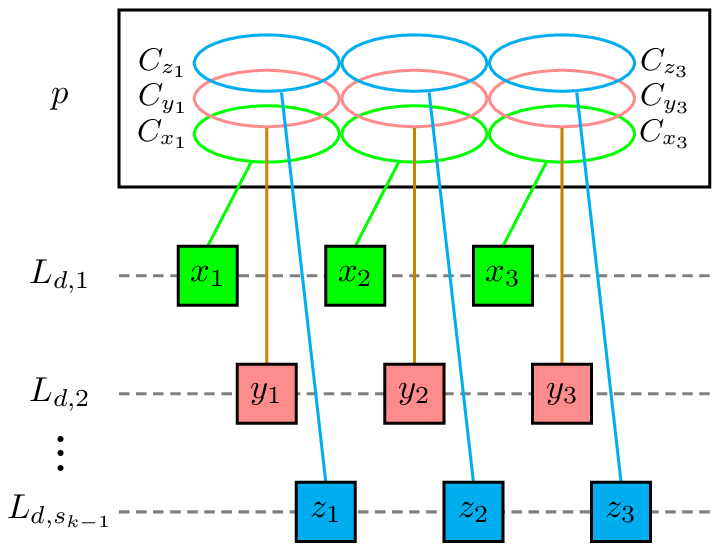}}{Track layout of nodes
with a common parent $p$.}

This construction defines a partial order on the nodes in track $L_{d,i}$, which can be arbitrarily extended to a total order. Hence we have a track assignment of $T$.  Since the nodes in each track are ordered primarily with respect to the relative positions of their parent nodes in the previous tracks, there is no X-crossing, and hence we have a track layout of $T$.  

%%%%%%%%%%%%%%%%%%%%%%%%%%%%%%%%%%%%%%%%%%%%%%%%%%
\subsubsection*{Track layout of {\boldmath $G$}}

To construct a track assignment of $G$ from the track layout of $T$, replace each track $L_{d,i}$ by $t_{k-1}$ `sub-tracks', and for each node $x$ of $T$, insert the track layout of $G[T_x]$ in place of $x$ on the sub-tracks corresponding to the track containing $x$ in the track layout of $T$.   More formally, the track layout of $G$ is indexed by the set 
\begin{equation*}
\{(d,i,j):d\geq0, 1\leq i\leq s_{k-1},1\leq j\leq t_{k-1}\}\enspace.
\end{equation*}
Each track  $V_{d,i,j}$ consists of those vertices $v$ of $G$ such that,  if $T_x$ is the bag containing $v$, then $x$ is at depth $d$ in $T$, $\alpha(x)=i$, and $v$ is in track $j$ in the track layout of $G[T_x]$. If $x$ and $y$ are distinct nodes of $T$ with $x<y$ in $L_{d,i}$, then  $v<w$ in $V_{d,i,j}$, for all vertices $v\in T_x$ and $w\in T_y$ in track $j$. If $v$ and $w$ are vertices of $G$ in track $j$  in bag $T_x$ at depth $d$,  then the relative order of $v$ and $w$ in $V_{d,\alpha(x),j}$ is the same as in the track layout of $G[T_x]$.

Clearly adjacent vertices of $G$ are in distinct tracks. Thus we have defined a track assignment of $G$. We claim there is no X-crossing. Clearly an intra-bag edge of $G$ is not in an X-crossing with an edge not in the same bag. By induction, there is no X-crossing between intra-bag edges in a common bag. Since there is no X-crossing in the track layout of $T$, inter-bag edges of $G$ which are mapped to edges of $T$ without a common parent node, are not involved in an X-crossing. 

Consider a parent node $p$ in $T$. For each child node $x$ of $p$, the set of vertices in $T_p$ adjacent to a vertex in $T_x$ forms the clique $C_x$. Thus there is no X-crossing between a pair of edges both from $C_x$ to $T_x$, since the vertices of $C_x$ are on distinct tracks.  Consider two child nodes $x$ and $y$ of $p$. For there to be an X-crossing  between an edge from $T_p$ to $T_x$ and an edge from $T_p$ to $T_y$, the nodes $x$ and $y$ must be on the same track in the track layout of $T$. Suppose $x<y$ in this track. By construction, $C_x$ and $C_y$ cover the same set of tracks, and $C_x\preceq C_y$ in the corresponding nice ordering. Thus for any track  containing vertices  $v\in C_x$ and $w\in C_y$, $v\leq w$ in that track. Since all the vertices in $T_x$ are to the left of the vertices in $T_y$ (in a common track), there is no X-crossing between an edge from $T_p$ to $T_x$ and an edge from $T_p$ to $T_y$. Therefore there is no X-crossing, and hence we have a track layout of $G$.

%%%%%%%%%%%%%%%%%%%%%%%%%%%%%%%%%%%%%%%%%%%%%%%%%%
\subsubsection*{Wrapped track layout of {\boldmath $G$}}

As illustrated in \figref{BigPicture}, we now `wrap' the track layout  of $G$ in the spirit of \lemref{TreeTrackLayout}. In particular, define a track assignment of $G$ indexed by 
\begin{equation*}
\big\{(d',i,j):d'\in\{0,1,2\},1\leq i\leq s_{k-1},1\leq j\leq t_{k-1}\big\}\enspace,
\end{equation*}
where each track 
\begin{equation*}
W_{d',i,j}\;=\;\bigcup\,\{V_{d,i,j}:d\equiv d'\pmod{3}\}\enspace.
\end{equation*}
If  $v\in V_{d,i,j}$ and $w\in V_{d+3,i,j}$ then $v<w$ in the order of $W_{d',i,j}$ (where $d'=d\bmod 3$). The order of each $V_{d,i,j}$ is preserved in $W_{d',i,j}$. The set of tracks $\{W_{d',i,j}:d'\in\{0,1,2\},1\leq i\leq s_{k-1},1\leq j\leq t_{k-1}\}$ forms a track assignment of $G$. 

For every edge $vw$ of $G$, the depths of the bags in $T$ containing $v$ and $w$ differ by at most one. Thus in the wrapped track assignment of $G$, adjacent vertices remain on distinct tracks, and there is no X-crossing. The number of tracks is  $3\cdot s_{k-1}\cdot t_{k-1}$.

Every clique $C$ of $G$ is either contained in a single bag of the tree-partition or is contained in two adjacent bags.  Let 
\begin{equation*}
\SSS'=\big\{\{(d',i,h):h\in X_j\}:d'\in\{0,1,2\},
1\leq i,j\leq s_{k-1}\big\}\enspace.
\end{equation*}
For every clique $C$ of $G$ contained in a single bag, the set of tracks containing $C$ is in $\SSS'$. Let 
\begin{align*}
\SSS''=\big\{
&\{(d',i,\ell):\ell\in X_j\}\cup
\{((d'+1)\bmod{3},p,h):h\in X_q\}:\\
&d'\in\{0,1,2\},1\leq i,j,p,q\leq s_{k-1}\big\}\enspace.
\end{align*}
For every clique $C$ of $G$ contained in two bags, the set of tracks containing
$C$ is in $\SSS''$. Observe that  $\SSS'\cup\SSS''$ is independent of $G$. Hence
$\SSS_k=\SSS'\cup\SSS''$ satisfies the hypothesis for $k$.

Now $|\SSS'|=3s_{k-1}^2$ and $|\SSS''|=3s_{k-1}^4$, and thus  $|\SSS'\cup\SSS''|=3s_{k-1}^2(s_{k-1}^2+1)$. Therefore any solution to the following set of recurrences  satisfies the theorem:  
\begin{equation}
s_0\;\geq\;1,\;\;\;\;
t_0\;\geq\;1,\;\;\;\;
s_k\;\geq\;3s_{k-1}^2(s_{k-1}^2+1),\;\;\;\;
t_k\;\geq\;3 s_{k-1}\cdot t_{k-1}\enspace.
\eqnlabel{MainRecurrence}
\end{equation}
We claim that 
\begin{equation*}
s_k=6^{(4^k-1)/3}\text{ and }t_k=3^k\cdot 6^{(4^k-3k-1)/9}
\end{equation*}
is a solution to \eqnref{MainRecurrence}. Observe that $s_0=1$ and $t_0=1$. Now
\begin{equation*}
3s_{k-1}^2(s_{k-1}^2+1)\;\leq\;6s_{k-1}^4\enspace,
\end{equation*}
and
\begin{equation*}
6(6^{(4^{k-1}-1)/3})^4\;=\; 6^{1+4(4^{k-1}-1)/3}
\;=\;6^{(4^k-1)/3}
\;=\;s_k\enspace.
\end{equation*}
Thus the recurrence for $s_k$ is satisfied. Now
\begin{align*}
3\cdot s_{k-1}\cdot t_{k-1}
\;=\;&3\cdot6^{(4^{k-1}-1)/3}\cdot3^{k-1}\cdot 6^{(4^{k-1}-3(k-1)-1)/9}\\
\;=\;&3^k\cdot6^{(3\cdot 4^{k-1}-3+4^{k-1}-3k+3-1)/9}\\
\;=\;&3^k\cdot6^{(4^k-3k-1)/9}\\
\;=\;&t_k\enspace.
\end{align*}
Thus the recurrence for $t_k$ is satisfied. This completes the proof.
\hfill\end{proof}

\Figure{BigPicture}{\includegraphics{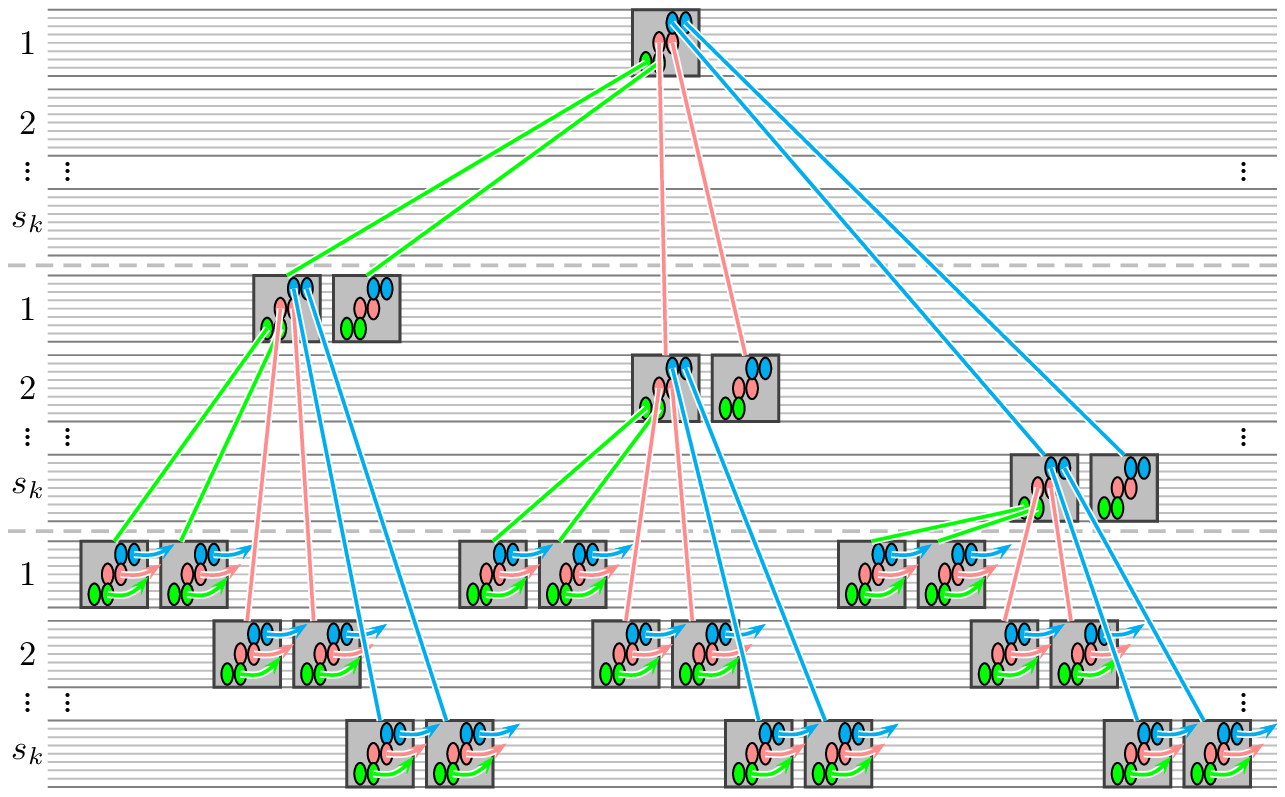}}{Wrapped track layout in \thmref{TreewidthTracknumber}.}

In the proof of \thmref{TreewidthTracknumber} we have made little effort to
reduce the bound on $t_k$, beyond that it is a doubly exponential function of
$k$. In  \cite{DujWoo-TR-02-03} we describe a number of refinements that result
in improved bounds on $t_k$. One such refinement uses strict $k$-trees. From an
algorithmic point of view, the disadvantage of using strict $k$-trees is that
at each recursive step, extra edges must be added to enlarge the graph from a
partial strict $k$-tree into a strict $k$-tree, whereas when using (non-strict)
$k$-trees, extra edges need only be added at the beginning of the algorithm.

For small values of $k$, much-improved results can be obtained. For example, we
prove that every series-parallel graph (that is, with tree-width at most two)
has an $18$-track layout \cite{DujWoo-TR-02-03}, whereas $t_2=54$. This bound
has recently been improved to $15$ by Di~Giacomo\etal\cite{DLM-TR03}. Their
method is based on  \twothmref{TreePartition}{TreewidthTracknumber}, and in the general case, still gives a doubly exponential upper bound on the track-number of graphs with tree-width $k$. For other particular classes of graphs, Di~Giacomo and Meijer~\cite{Giacomo-GD03,DM-GD03} recently improved the constants in our results.

Our doubly exponential upper bound is probably not best possible. Di~Giacomo\etal\cite{DLM-TR03} constructed graphs with tree-width $k$ and track-number at least $2k+1$. The following construction establishes a quadratic lower bound. It is similar to a graph due to Albertson~\cite{Albertson-EJC04}, which gives a tight lower bound on the star chromatic number of graphs with tree-width $k$.

\begin{theorem}
\thmlabel{TreewidthLowerBound}
For all $k\geq0$, there is a graph $G_k$ with tree-width at most $k$ and track-number $\tn{G_k}=\half(k+1)(k+2)$.
\end{theorem}

\begin{proof}
Let $G_0=K_1$. Obviously $G_0$ has tree-width $0$. Construct $G_k$ from $G_{k-1}$ as follows.  Start with a $k$-clique $\{v_1,v_2,\dots,v_k\}$. Let $n=2(\half(k+1)(k+2)-1-k)+1$. Add $n$ vertices $\{w_1,w_2,\dots,w_n\}$ each adjacent to every $v_i$. Let $H_1,H_2,\dots,H_n$ be copies of $G_{k-1}$. For all $1\leq j\leq n$, add an edge between $w_j$ and each vertex of $H_j$. It is easily seen that from a tree decomposition of $G_{k-1}$ of width $k-1$, we can construct a tree decomposition of $G_k$ of width $k$. Thus $G_k$ has tree-width at most $k$.

\Figure{TreewidthConstruction}{\includegraphics{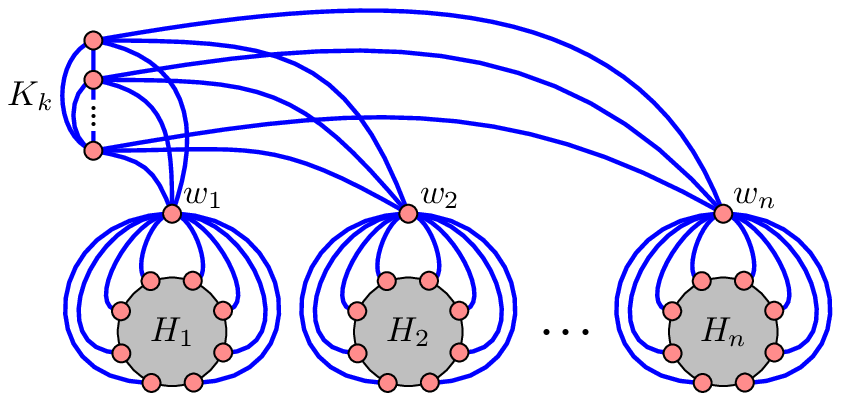}}{The graph $G_k$.}

To prove that $\tn{G_k}\geq\half(k+1)(k+2)$, we proceed by induction on $k\geq0$. Obviously $\tn{G_0}=1$. Suppose that $\tn{G_{k-1}}\geq\half k(k+1)$, but $\tn{G_k}\leq\half(k+1)(k+2)-1$. Since $\{v_1,v_2,\dots,v_k\}$ is a clique,  we can assume that $v_i$ is in track $i$. Since each vertex $w_j$ is adjacent to each $v_i$, no $w_j$ is in tracks $\{1,2,\dots,k\}$. There are $\half(k+1)(k+2)-1-k$ remaining tracks.  Since $n$ is more than twice this number, there are at least three $w_j$ vertices in a single track.  Without loss of generality, $w_1<w_2<w_3$ in track $k+1$. No vertex $x$ of $H_2$ is in track $i\in\{1,2,\dots,k\}$, as otherwise  $xw_2$ would form an X-crossing with $v_iw_1$ or $v_iw_3$. No vertex $x$ of $H_2$ is in track $k+1$, since $x$ and $w_2$ are adjacent, and $w_2$ is in track $k+1$. Thus all the vertices of $H_2$ are in tracks $\{k+2,k+3,\dots,\half(k+1)(k+2)-1\}$.  There are $\half(k+1)(k+2)-1-(k+1)=\half k(k+1)-1$ such tracks.  This contradicts the assumption that  $\tn{G_{k-1}}\geq\half k(k+1)$. Therefore $\tn{G_k}\geq\half(k+1)(k+2)$.

It remains to prove that $\tn{G_k}\leq\half(k+1)(k+2)$. Suppose we have a $\half k(k+1)$-track layout of $G_{k-1}$. Thus each $H_j$ has a $\half k(k+1)$-track layout. Put each vertex $v_i$ of $G_k$ in track $i$. Put the vertices $\{w_1,w_2,\dots,w_n\}$ in track $k+1$ in this order. Put the track layout of each $H_j$ in tracks $k+2,k+3,\dots,\half(k+1)(k+2)$, such that the vertices of $H_j$ precede the vertices of $H_{j+1}$. Clearly there are no X-crossings.
\hfill\end{proof}

Also note that \thmref{TreewidthLowerBound} (for $k\geq2$) can be extended using the proof technique of \lemref{AddingEars} to give the same lower bound for improper track layouts.

%%%%%%%%%%%%%%%%%%%%%%%%%%%%%%%%%%%%%%%%%%%%%%%%%%%%%%%%%%%%%%%%%%%%%%%%%%%%%%
\section{Open Problems}
\seclabel{Conclusion}
%%%%%%%%%%%%%%%%%%%%%%%%%%%%%%%%%%%%%%%%%%%%%%%%%%%%%%%%%%%%%%%%%%%%%%%%%%%%%%%

\begin{remunerate}

\item (In the conference version of their paper) Felsner~\cite{FLW-JGAA03} asked whether every planar graph has  a three-dimensional drawing with \Oh{n} volume? By \thmref{DrawingIFFTracknumber}, this question has an affirmative answer if every planar graph has \Oh{1} track-number. Whether every planar graph has \Oh{1} track-number is an open problem due to H.\ de Fraysseix [private communication, 2000], and by \thmref{QueuenumberIFFTracknumber}, is equivalent to the following question.

\item Heath\etal\cite{HR-SJC92,HLR-SJDM92} asked whether every planar graph has \Oh{1} queue-number? The best known upper bound on the queue-number of a planar graph is $\Oh{\sqrt{n}}$. In general, Dujmovi{\'c} and Wood~\cite{DujWoo-LinearLayouts} proved that every $m$-edge graph has queue-number at most $\e\sqrt{m}$, where $\e$ is the base of the natural logarithm. 

\item Heath\etal\cite{HR-SJC92,HLR-SJDM92} asked whether stack-number is bounded by queue-number (and vice-versa)? Note that there is a family of graphs $\G$ with $\sn{G}\in\Omega(3^{\Omega(\qn{G})-\epsilon})$, for all $G\in\G$ \cite{HLR-SJDM92}.

\item Is the queue-number of a graph bounded by  a polynomial (or even singly
exponential) function of its tree-width? 

\end{remunerate}

%%%%%%%%%%%%%%%%%%%%%%%%%%%%%%%%%%%%%%%%%%%%%%%%%%%%%%%%%%%%%%%%%%%%%%%%%%%%%%%%
\section*{Acknowledgements} 

The authors are grateful for stimulating discussions with Prosenjit Bose, Jurek Czyzowicz, Hubert de Fraysseix, Stefan Langerman, Giuseppe Liotta, Patrice Ossona de Mendez, and Matthew Suderman. Thanks to an anonymous referee for many helpful comments.

%a beautiful f.\ review.

%%%%%%%%%%%%%%%%%%%%%%%%%%%%%%%%%%%%%%%%%%%%%%%%%%%%%%%%%%%%%%%%%%%%%%%%%%%%%%%%
\bibliographystyle{siam}%{myBibliographyStyle}
\bibliography{myBibliography,myConferences}
%%%%%%%%%%%%%%%%%%%%%%%%%%%%%%%%%%%%%%%%%%%%%%%%%%%%%%%%%%%%%%%%%%%%%%%%%%%%%%%%

\end{document}